\documentclass[11pt]{article}
\usepackage{graphicx,amsmath,amssymb,float}
\usepackage{subfigure}
\usepackage{multirow,tabu,makecell}

\begin{document}

\newtheorem{theorem}{Theorem}[section]
\newtheorem{lemma}[theorem]{Lemma}
\newtheorem{corollary}[theorem]{Corollary}
\newtheorem{proposition}[theorem]{Proposition}
\newcommand{\blackslug}{\penalty 1000\hbox{
    \vrule height 8pt width .4pt\hskip -.4pt
    \vbox{\hrule width 8pt height .4pt\vskip -.4pt
          \vskip 8pt
      \vskip -.4pt\hrule width 8pt height .4pt}
    \hskip -3.9pt
    \vrule height 8pt width .4pt}}
\newcommand{\proofend}{\quad\blackslug}
\newenvironment{proof}{$\;$\newline \noindent {\sc Proof.}$\;\;\;$\rm}{\qed}
\newcommand{\qed}{\hspace*{\fill}\blackslug}
\newenvironment{definition}{$\;$\newline \noindent {\bf Definition}$\;$}{$\;$\newline}
\def\boxit#1{\vbox{\hrule\hbox{\vrule\kern4pt
  \vbox{\kern1pt#1\kern1pt}
\kern2pt\vrule}\hrule}}
\addtolength{\baselineskip}{+0.4mm}

\title{Algorithms for Maximum Agreement Forest\\ of Multiple General Trees}
\author{
 \vspace*{3mm}
 {\sc Feng Shi}$^{\mbox{\footnotesize \textdagger}}$ \ \ 
 {\sc Jianer Chen}$^{\mbox{\footnotesize \textdagger\textdaggerdbl}}$ \ \
 {\sc Qilong Feng}$^{\mbox{\footnotesize \textdagger}}$ \ \ \\
 {\sc Xiaojun Ding}$^{\mbox{\footnotesize \textdagger}}$ \ \vspace*{3mm}
 {\sc Jianxin Wang}$^{\mbox{\footnotesize \textdagger}}$\\ \
   $^{\mbox{\footnotesize \textdagger}}$School of Information Science and Engineering \\
   Central South University \\
  \vspace*{3mm}
   Changsha 410083, P.R. China\\
   $^{\mbox{\footnotesize \textdaggerdbl}}$Department of Computer Science and Engineering\\
   Texas A\&M University\\
   College Station, Texas 77843-3112, USA  }

\date{}
\maketitle

\vspace{-7mm}

\begin{abstract}
The Maximum Agreement Forest ({\sc Maf}) problem is a well-studied problem in evolutionary biology, which asks for a largest common subforest of a given collection of phylogenetic trees with identical leaf label-set. However, the previous work about the {\sc Maf} problem are mainly on two binary phylogenetic trees or two general (i.e., binary and non-binary) phylogenetic trees. In this paper, we study the more general version of the problem: the {\sc Maf} problem on multiple general phylogenetic trees. We present a parameterized algorithm of running time $O(3^k n^2m)$ and a $3$-approximation algorithm for the {\sc Maf} problem on multiple rooted general phylogenetic trees, and a parameterized algorithm of running time $O(4^k n^2m)$ and a $4$-approximation algorithm for the {\sc Maf} problem on multiple unrooted general phylogenetic trees. We also implement the parameterized algorithm and approximation algorithm for the {\sc Maf} problem on multiple rooted general phylogenetic trees, and test them on simulated data and biological data.
\end{abstract}

\section{Introduction}

Phylogenetic trees (evolutionary trees) are widely used in evolutionary biology to represent the tree-like evolution of a collection of extant species. However, due to reticulation events, such as hybridization and lateral gene transfer (LGT) in evolution, phylogenetic trees representing the evolutionary history of different parts of the genomes found in the same collection of the extant species may differ. In order to facilitate the comparison of these different phylogenetic trees, several metrics were proposed in the literature, such as Robinson-Foulds distance \cite{1}, NNI (Nearest Neighbor Interchange) distance \cite{2}, TBR (Tree Bisection and Reconnection) distance, SPR (Subtree Prune and Regraft) distance \cite{3,4}, and Hybridization number \cite{hybridization}.

The SPR distance between two phylogenetic trees is the minimum number of 'subtree prune and regraft' operations \cite{6} that needed to  convert one tree to the other, which is equivalent to the minimum number of reticulation events to transform between the two trees. Thus, the SPR distance provides a lower bound on the number of such events needed to reconcile the two phylogenetic trees \cite{10}. And this lower bound gives an indication of the extent to which reticulation has influenced the evolutionary history of the extant species under consideration. Similarly to the definition of the SPR distance, the TBR distance between two phylogenetic trees is the minimum number of `tree bisection and reconnection' operations \cite{6} that needed to convert one tree to the other. Although the TBR distance has no known direct biological meaning, it can be used to bound the space of phylogenetic trees.

For the study of TBR distance and SPR distance, a graph theoretical model, the {\it maximum agreement forest} (MAF) of two phylogenetic trees, has been formulated. Define the {\it order} of a forest to be the number of connected components in the forest.\footnote{The definitions for the study of maximum agreement forests have been kind of confusing. If {\it size} denotes the number of edges in a forest, then for a forest, the size is equal to the number of vertices minus the order. In particular, when the number of vertices is fixed, a forest of a large size means a small order of the forest.} Allen and Steel \cite{6} proved that the TBR distance between two {\it unrooted binary} phylogenetic trees is equal to the order of their MAF minus $1$, and Bordewich and Semple \cite{7} proved that the rSPR distance between two {\it rooted binary} phylogenetic trees is equal to the order of their rooted version of MAF minus $1$. Therefore, there are extensive researches studying the {\sc Maf} problem, which asks for constructing an MAF for the given two phylogenetic trees.

Traditionally, biological researchers assumed that phylogenetic trees are bifurcating \cite{21,22}, which leads to most previously work about the {\sc Maf} problem are restricted to binary trees. However, for many biological data sets in practice (e.g., \cite{27,28}), the constructed phylogenetic trees are not strictly bifurcating, that is, these trees contain polytomies. There are two reasons for the polytomies in phylogenetic trees. First, lack of sufficient data or inappropriate analysis of characters, which result in poor resolution of true bifurcating relationships. Second, multiple, simultaneous speciation events \cite{23}. Moreover, more recent evidences show the existence of simultaneous speciation events (e.g., \cite{24,25,26}). Therefore, it is crucial to study the general (i.e., binary and non-binary) phylogenetic trees. Note that, it is not difficult to verify that the order of an MAF for two rooted general trees minus 1 is equal to their SPR distance, and the order of an MAF for two unrooted general trees minus 1 is equal to their TBR distance.

Note again that we may construct multiple (i.e., two or more) different phylogenetic trees for the same collection of species according to different data sets or different building methods. Constructing an MAF for these different trees makes more biological meaning than just for two trees. There are two reasons. First, take the MAF for two rooted phylogenetic trees for instance, we have mentioned above that the SPR distance between two trees provides a lower bound on the number of reticulation events needed to reconcile the two trees. But these two trees only represent the evolutionary histories of two different parts of the genomes found in the collection of species, thus, this lower bound can only give an indication of the extent to which reticulation has influenced the evolutionary histories of the two parts of the genomes found in the collection of species. If we construct a phylogenetic tree for each part of the genomes found in the collection of species and compare these different trees simultaneously, then, similarly, we can also have the same conclusion that the order of the MAF for these trees provides a lower bound on the number of reticulation events needed to reconcile these trees. And this lower bound can give a more comprehensive indication of the extent to which reticulation has influenced the evolutionary history of the collection of species. Second, constructing the MAF for multiple trees is a critical step in studying the reticulate networks of multiple phylogenetic trees \cite{18}, which is a hot issue in phylogenetics. Therefore, it is meaningful to study the {\sc Maf} problem on multiple trees. However, the {\sc Maf} problem on multiple trees has not been studied as extensively as that on two trees.

Above all, it makes perfect sense to investigate the {\sc Maf} problem on multiple general phylogenetic trees. In this paper, we will focus on the approximation algorithms and parameterized algorithms for the {\sc Maf} problem on multiple rooted general trees, and for the {\sc Maf} problem on multiple unrooted general trees.

In the following, we firstly review the previous related work about the {\sc Maf} problem. In terms of computational complexity, it is known that computing the order of an MAF is NP-hard and MAX SNP-hard for two unrooted binary phylogenetic trees \cite{5}, as well as for two rooted binary phylogenetic trees \cite{7}.

\smallskip

{\bf Approximation Algorithms}. For the {\sc Maf} problem on two rooted binary phylogenetic trees, Hein et al.\;\cite{5} proposed an approximation algorithm of ratio $3$. But Rodrigues et al.\;\cite{8} found a subtle error in \cite{5}, showed that the algorithm in \cite{5} has ratio at least $4$, and presented a new approximation algorithm which they claimed has ratio $3$. Bonet et al.\;\cite{9} provided a counterexample and showed that both the algorithms in \cite{5} and \cite{8} compute a $5$-approximation of the rSPR distance between two rooted binary trees in linear time. The approximation ratio was improved to 3 by Bordewich et al.\;\cite{10}, but at the expense of an increased running time of $O(n^5)$. A second 3-approximation algorithm presented in \cite{11} achieves a running time of $O(n^2)$. Whidden et al.\;\cite{12,13} presented the third 3-approximation algorithm, which runs in linear-time. Recently, Shi et al.\;\cite{shiapprox} presented an improved approximation algorithm of ratio $2.5$, which is the best known approximation algorithm for the {\sc Maf} problem on two rooted binary trees. For the {\sc Maf} problem on two unrooted binary phylogenetic trees, Whidden et al.\;\cite{12,13} presented a linear-time best known approximation algorithm of ratio $3$.

There is also a couple of approximation algorithms for the {\sc Maf} problem on two general phylogenetic trees. Rodrigues et al.\;\cite{11} developed an approximation algorithm of ratio $d+1$ for the {\sc Maf} problem on two rooted general trees, where $d$ is the maximum number of children a node in the input trees may have. Chen et al.\;\cite{14} developed a $3$-approximation algorithm, which is the first constant-ratio approximation algorithm for the {\sc Maf} problem on two unrooted general trees.

For the {\sc Maf} problem on multiple rooted binary phylogenetic trees, Chataigner \cite{15} presented an $8$-approximation algorithm. Recently, the approximation ratio was improved to $3$ by Shi et al.~\cite{20}. For the {\sc Maf} problem on multiple unrooted binary trees, Shi et al.~\cite{20} presented a $4$-approximation algorithm. To our best knowledge, there is currently no known approximation algorithm for the {\sc Maf} problem on multiple rooted (unrooted) general phylogenetic trees.

\smallskip

{\bf Parameterized Algorithms}. Parameterized algorithms for the {\sc Maf} problem, parameterized by the order $k$ of an MAF, have also been studied. A parameterized problem is {\it fixed-parameter tractable} \cite{fptbook} if it is solvable in time $f(k)n^{O(1)}$, where $k$ is the parameter and $n$ is the input size. For the {\sc Maf} problem on two unrooted binary phylogenetic trees, Allen and Steel \cite{6} showed that is fixed-parameter tractable. Hallett and McCartin \cite{10} developed a faster parameterized algorithm of running time $O(4^k k^5+n^{O(1)})$. Whidden and Zeh \cite{11} further improved the time complexity to $O(4^k k + n^3)$ or $O(4^k n)$. For the {\sc Maf} problem on two rooted binary phylogenetic trees, Bordewich {\it et al.}~\cite{10} developed a parameterized algorithm of running time $O(4^k k^4 + n^3)$. Whidden {\it et al.}~\cite{12,13} improved this bound and developed an algorithm of running time $O(2.42^k k + n^3)$. Chen {\it et al.}~\cite{zhizhongchen} presented currently the fastest algorithm of running time $O(2.344^k n)$ for the {\sc Maf} problem on two rooted binary trees.

There is also a couple of parameterized algorithms for the {\sc Maf} problem on two general phylogenetic trees. Whidden {\it et al.}~\cite{17} presented a parameterized algorithm of running time $O(2.42^k k + n^3)$ for the {\sc Maf} problem on two rooted general trees. And Chen et al.\;\cite{14} developed an algorithm of running time $O(3^k n)$ for the {\sc Maf} problem on two unrooted general trees, which is also currently the fastest algorithm for the {\sc Maf} problem on two unrooted binary trees.

For the {\sc Maf} problem on multiple rooted binary phylogenetic trees, Chen et al.~\cite{18} presented a parameterized algorithm of running time $O^*(6^k)$. Shi et al.~\cite{19} improved this bound and developed an algorithm of running time $O(3^k n)$. For the {\sc Maf} problem on multiple unrooted binary phylogenetic trees, Shi et al.~\cite{19} presented the first parameterized algorithm of running time $O(4^k n)$. To our best knowledge, there is currently no known parameterized algorithm for the {\sc Maf} problem on multiple rooted (unrooted) general phylogenetic trees.

\smallskip

{\bf Contributions}. In the current paper, we are focused on the approximation algorithms and parameterized algorithms for the {\sc Maf} problem on multiple general phylogenetic trees, for both the version of rooted trees and the version of unrooted trees. Our algorithms are based on careful analysis of the graph structures that takes advantage of special relations among leaves in the trees.
Our main contributions include two parameterized algorithms for the Maximum Agreement Forest problem on multiple general trees: one for rooted trees that runs in time $O(3^k n^2m)$, and the other for unrooted trees that runs in time $O(4^k n^2m)$. And two approximation algorithms for the Maximum Agreement Forest problem on multiple general trees: one for rooted trees with ratio $3$, and the other for unrooted trees with ratio $4$.

We implement the approximation algorithm and parameterized algorithm for the {\sc Maf} problem on multiple rooted general phylogenetic trees, obtain programs {\sc Amaf} and {\sc Pmaf}, respectively. We test both programs on simulated data and biological data. Given multiple rooted general trees, {\sc Pmaf} can calculate the order of an MAF for these trees rapidly when the order of an MAF is small. And the order of the agreement forest returned by {\sc Amaf} is always less than $3$ times the order of an MAF for these trees.

\section{Definitions and Problem Formulations}

A tree is a {\it single-vertex tree} if it consists of a single vertex, which is the leaf of the tree. A tree is {\it general} if either it is a single-vertex tree or each of its vertices has degree either $1$ or greater than $2$. The degree-$1$ vertices are {\it leaves} and the other vertices are {\it non-leaves} of the tree. There are two kinds of trees in our discussion, one is unrooted trees and the other is rooted trees. In the following, we first give the terminologies on the unrooted trees, then remark on the differences for the rooted trees. Let $X$ be a fixed irrelevant {\it label-set}.

\subsection{Unrooted $X$-trees and $X$-forests}


A general tree is {\it unrooted} if no root is specified in the tree -- in this case no ancestor-descendant relation is defined in the tree. For the label-set $X$, an unrooted {\it general phylogenetic $X$-tree}, or simply an unrooted {\it $X$-tree}, is an unrooted general tree whose leaves are labeled bijectively by the label-set $X$ (all non-leaves are unlabeled). A {\it subforest} of an unrooted $X$-tree $T$ is a subgraph of $T$. And a {\it subtree} $T'$ of $T$ is a connected subgraph of $T$, which contains at least one leaf in $T$. Denote by $L(T')$ the label set that contains all labels in $T'$. An unrooted {\it $X$-forest $F$} is a subforest of an unrooted $X$-tree $T$ that contains all leaves of $T$ such that each connected component of $F$ contains at least one leaf in $T$. Thus, an unrooted $X$-forest $F$ is a collection of subtrees of $T$, moreover, the label-sets of these subtrees are disjoint and the union of the label-sets is equal to $X$. Define the {\it order} of the $X$-forest $F$, denoted Ord$(F)$, to be the number of connected components in $F$.

A subtree $T'$ of an unrooted $X$-tree may contain unlabeled vertices of degree less than $3$. In this case we apply the {\it forced contraction} operation on $T'$, which replaces each degree-$2$ vertex $v$ and its incident edges with a single edge connecting the two neighbors of $v$, and removes each unlabeled vertex that has degree smaller than $2$. Note that the forced contraction does not change the order of an $X$-forest. It has been well-known that the forced
contraction operation does not affect the construction of an MAF for $X$-trees. Therefore, we will assume that the forced contraction is applied immediately whenever it is applicable. An $X$-forest $F$ is {\it irreducible} if the forced contraction can not apply to $F$. Thus, the $X$-forests in our discussion are always assumed to be irreducible. With this assumption, each unlabeled vertex in an unrooted $X$-forest has degree not less than $3$.

Two $X$-forests $F_1$ and $F_2$ are isomorphic if there is a graph isomorphism between $F_1$ and $F_2$ in which each leaf of $F_1$
is mapped to a leaf of $F_2$ with the same label. We will simply say that an $X$-forest $F'$ is a subforest of another $X$-forest $F$ if, up to the forced contraction, $F'$ is isomorphic to a subforest of $F$.

\subsection{Rooted $X$-trees and $X$-forests}


A general tree is {\it rooted} if a particular leaf is designated as the root (so it is {\it both} a root and a leaf), which specifies a unique ancestor-descendant relation in the tree. A rooted $X$-{\it tree} is a rooted general tree whose leaves are labeled bijectively by the label-set $X$. The root of a rooted $X$-tree will always be labeled by a special label $\rho$, which is always assumed to be in the label-set $X$. A subtree $T'$ of a rooted $X$-tree $T$ is a connected subgraph of $T$ which contains at least one leaf in $T$. In order to preserve the ancestor-descendant relation in $T$, we should define the root of the subtree $T'$. If $T'$ contains the leaf labeled $\rho$, certainly, it is the root of the subtree; otherwise, the node in $T'$ that is in $T$ the least common ancestor of all the labeled leaves in $T'$ is defined to be the root of $T'$. A {\it subforest} of a rooted $X$-tree $T$ is defined to be a subgraph of $T$. A rooted {\it $X$-forest $F$} is a subforest of a rooted $X$-tree $T$ that contains a collection of subtrees whose label-sets are disjoint such that the union of the label-sets is equal to $X$. Thus, one of the subtrees in a rooted $X$-forest $F$ must have the leaf labeled $\rho$ as its root.

We also assume that the forced contraction is applied immediately whenever it is applicable. However, if the root $r$ of a subtree $T'$ is of degree $2$, then the operation will {\it not} be applied on $r$, in order to preserve the ancestor-descendant relation in $T'$. Thus, all unlabeled vertices in $T'$ that are not the root of $T'$ have degree not less than $3$.

\subsection{Agreement Forest}


The following terminologies are used for both rooted trees and unrooted trees.

An $X$-forest $F$ is an {\it agreement forest} for a collection $\{F_1,F_2,\ldots,F_m\}$ of $X$-forests if $F$ is a subforest of $F_i$, for all $i$. A {\it maximum agreement forest} (abbr. MAF) $F^*$ for $\{F_1, F_2, \ldots, F_m\}$ is an agreement forest for $\{F_1, F_2, \ldots, F_m\}$ with a minimum Ord$(F^*)$ over all agreement forests for $\{F_1, F_2, \ldots, F_m\}$.

The four versions of the {\sc Maf} problem on multiple general $X$-forests studied in the current paper, are formally given as follows.

\vspace*{-1mm}

\begin{quote}
{\sc rooted parameterized maximum agreement forest} ({\sc para-rMaf})\\
{\it Input}: \ \ A set $\{F_1, \ldots, F_m\}$ of rooted general $X$-forests, and a parameter $k$\\
{\it Output}: an agreement forest $F^*$ for $\{F_1, \ldots, F_m\}$ with Ord$(F^*) \leq k$\\
\hspace*{14mm}or report that no such an agreement forest exists\\

\vspace*{-1mm}

{\sc unrooted parameterized maximum agreement forest} ({\sc para-uMaf})\\
{\it Input}: \ \ A set $\{F_1, \ldots, F_m\}$ of unrooted general $X$-forests, and a parameter $k$\\
{\it Output}: an agreement forest $F^*$ for $\{F_1, \ldots, F_m\}$ with Ord$(F^*) \leq k$\\
\hspace*{14mm}or report that no such an agreement forest exists\\

\vspace*{-1mm}

{\sc rooted maximum agreement forest} ({\sc app-rMaf})\\
{\it Input}: \ \ A set $\{F_1, \ldots, F_m\}$ of rooted general $X$-forests\\
{\it Output}: a maximum agreement forest $F^*$
     for $\{F_1, \ldots, F_m\}$ \\

\vspace*{-1mm}

{\sc unrooted maximum agreement forest} ({\sc app-uMaf})\\
{\it Input}: \ \ A set $\{F_1, \ldots, F_m\}$ of unrooted general $X$-forests\\
{\it Output}: a maximum agreement forest $F^*$
     for $\{F_1, \ldots, F_m\}$
\end{quote}

Every agreement forest $F$ for a collection $\{F_1,F_2,\ldots,F_m\}$ of $X$-forests corresponds to a unique minimum subgraph (contains the minimum number of edges) of $F_i$, denoted by $F_i^{F}$, for all $i$. Thus, without any confusion, we can simply say that an edge $e$ is in or not in the agreement forest $F$, as long as $e$ is in or not in the unique corresponding subgraph $F_i^{F}$, respectively.

The following concept on two $X$-forests will play an important role in our discussion.

\begin{definition}
Let $F_1$ and $F_2$ be two $X$-forests (either both rooted or both unrooted). An agreement forest $F$ for $F_1$ and $F_2$ is a {\it maximal agreement forest} ({\it maximal-AF}) for $F_1$ and $F_2$ if there is no agreement forest $F'$ for $F_1$ and $F_2$ such that $F$ is a subforest of $F'$ and $\mbox{Ord$(F')$} < \mbox{Ord$(F)$}$.
\end{definition}

By definition, an MAF for two $X$-forests $F_1$ and $F_2$ is also a maximal-AF for $F_1$ and $F_2$, but the inverse is not necessarily true.

\section{Reduction Rule for {\sc Maf}}

Fix a label-set $X$. Because of the bijection between the leaves in an $X$-forest $F$ (either rooted or unrooted) and the labels in the label-set $X$, sometimes we will use, without confusion, a label in $X$ to refer to the corresponding leaf in $F$, or vice versa.

For a subset $E'$ of edges in an $X$-forest $F$ (either rooted or unrooted), we will denote by $F \setminus E'$ the forest $F$ with the edges in $E'$ removed. For any $X$-forest $F'$ that is a subforest of $F$, it is easy to see that there is an edge subset $E$ of $F$ that $F' = F \setminus E$.

For an arbitrary edge $e$ in an $X$-forest $F$, removing edge $e$ would lead two new subtrees be constructed in $F \setminus \{e\}$, denoted by $T_e^1$ and $T_e^2$, respectively.

Let $\{F_1,F_2,\ldots,F_m\}$ be a collection of $X$-forests (either all are rooted or all are unrooted), $m \geq 2$. In the following, we give a reduction rule for $\{F_1,F_2,\ldots,F_m\}$.

\smallskip

\noindent{\bf Reduction Rule 1.} Let $T_1,\ldots,T_t$ be several subtrees in $X$-forest $F_p$, $t \geq 1$, $1 \leq p \leq m$. If there exists an edge $e$ in $X$-forest $F_q$, $p \neq q$, $1 \leq q \leq m$, that $L(T_e^1) \subseteq (L(T_1) \cup \ldots \cup L(T_t))$ and $L(T_e^2) \cap (L(T_1) \cup \ldots \cup L(T_t)) = \emptyset$, then remove $e$ from $F_q$.

\begin{lemma}
\label{Reduction Rule 1}
Let $\{F'_1,F'_2,\ldots,F'_m\}$ be the collection that produced by Reduction Rule 1 on the collection $\{F_1,F_2,\ldots,F_m\}$, then $\{F'_1,F'_2,\ldots,F'_m\}$ and $\{F_1,F_2,\ldots,F_m\}$ have the same collection of MAFs.

\begin{proof}
Let $F$ be a fixed MAF for $\{F_1,F_2,\ldots,F_m\}$. Let $Y= L(T_1) \cup \ldots \cup L(T_t)$ and $Y'= X \setminus Y$. Since $F$ is a subforest of $F_p$, for each subtree $T_i$ in $F_p$, $1 \leq i \leq t$, we have that any label of $L(T_i)$ cannot be in the same connected component with any label of $X \setminus L(T_i)$ in $F$. Thus, any label of $Y$ cannot be in the same connected component with any label of $Y'$ in $F$.

Suppose edge $e$ is in $F$. Then there would exist a path in $F$ that connects a label of $L(T_e^1)$ and a label of $L(T_e^2)$. Because $L(T_e^1) \subseteq Y$ and $L(T_e^2) \subseteq Y'$, so there would exist a path in $F$ that connects a label of $Y$ and a label of $Y'$, which contracts the fact that any label of $Y$ cannot be in the same connected component with any label of $Y'$ in $F$. Thus, edge $e$ could not be in $F$. Therefore, $F$ is still a subforest of $F_q \setminus \{e\}$, and $F$ is also an MAF for $\{F'_1,F'_2,\ldots,F'_m\}$.
\end{proof}
\end{lemma}

We will assume that Reduction Rule 1 is applied whenever it is possible. A instance (of anyone of the four versions of {\sc Maf} problem) is {\it strongly reducible} if Reduction Rule 1 is not applicable on it. Therefore, all instances in our following discussion are always strongly reducible.

\section{Parameterized Algorithms}

Before analyzing the detail parameterized algorithms for {\sc para-rMaf} and {\sc para-uMaf}, we firstly give a few lemmas, which hold true for both {\sc para-rMaf} and {\sc para-uMaf}. And according to these lemmas, we can present the general frame of our parameterized algorithms.

The first lemma follows directly from the definition of maximal-AF.

\begin{lemma}
\label{maximal}
Every agreement forest for two $X$-forests $F_1$ and $F_2$ is a subforest of a maximal-AF $F'$ for $F_1$ and $F_2$, but $F'$ may not be unique.
\end{lemma}

Since the MAF $F$ for a set of $X$-forests $\{F_1,F_2,\ldots,F_m\}$ must be an agreement forest for $F_1$ and $F_2$, thus, by Lemma~\ref{maximal}, there must exist a maximal-AF $F'$ for $F_1$ and $F_2$ that $F$ is a subforest of $F'$. Moreover, we have the following lemma.

\begin{lemma}
\label{maximal-MAF}
Let $\{F_1,F_2,F_3,\ldots,F_m\}$ be a set of $X$-forests, and let $F$ be a fixed MAF for it. There must exist a maximal-AF $F'$ for $F_1$ and $F_2$ that $F$ is also an MAF for $\{F',F_3,\ldots,F_m\}$.

\begin{proof}
Let $F$ be a fixed MAF for $\{F_1, F_2, F_3, \ldots, F_m\}$, and let $F'$ be a maximal-AF for $F_1$ and $F_2$ that $F$ is a subforest of $F'$. Obviously, $F$ is an agreement forest for  $\{F',F_3,\ldots,F_m\}$, thus, the order of the MAF for $\{F',F_3,\ldots,F_m\}$ is not larger than Ord$(F)$. On the other hand, every agreement forest for $\{F',F_3,\ldots,F_m\}$ is also an agreement forest for $\{F_1,F_2,F_3,\ldots,F_m\}$, thus, the order of the MAF for $\{F',F_3,\ldots,F_m\}$ is not less than Ord$(F)$. Therefore, the order of the MAF for $\{F',F_3,\ldots,F_m\}$ is Ord$(F)$, $F$ is an MAF for $\{F',F_3,\ldots,F_m\}$.
\end{proof}
\end{lemma}

Let $(F_1,F_2,F_3,\ldots,F_m; k)$ be an instance of either {\sc para-rMaf} or {\sc para-uMaf}. Now according to Lemma~\ref{maximal-MAF}, we can give the general frame of our parameterized algorithms.

\begin{quote}
Main-Algorithm \\
1.  construct a collection ${\cal C}$ of agreement forests for $F_1$ and $F_2$ that \\
\hspace*{5mm}contains all maximal-AF $F$ for $F_1$ and $F_2$ with Ord$(F) \leq k$;\\
2.  {\bf for} each $F$ in the collection $\cal C$ constructed in step 1\\
\hspace*{4mm} {\bf do} recursively work on the instance $(F, F_3, \ldots, F_m; k)$.
\end{quote}

For an $X$-subforest $F'$ of an $X$-forest $F$, we always have Ord$(F') > \mbox{Ord}(F)$. Thus, no maximal-AF $F$ for $F_1$ and $F_2$ with Ord$(F) > k$ can contain an MAF $F'$ for $(F_1,F_2,F_3,\ldots,F_m; k)$ with Ord$(F') \leq k$. Therefore, in Step 1 of Main-Algorithm, we only need examine all maximal-AFs whose order is bounded by $k$.

\begin{theorem}
\label{main-para}
The Main-Algorithm correctly returns an agreement forest $F^*$ with Ord$(F^*) \leq k$ for $(F_1,F_2,F_3,\ldots,F_m;k)$ if such an agreement forest exists.

\begin{proof}
If there exists an agreement forest $F^*$ with Ord$(F^*) \leq k$ for $(F_1,F_2,F_3,\ldots,F_m;k)$, by Lemma~\ref{maximal-MAF}, there must exist a maximal-AF $F$ for $F_1$ and $F_2$ that $F^*$ is also an MAF for $(F,F_3,\ldots,F_m)$, which is an instance examined in Step 2. Therefore, if $(F_1,F_2,F_3,\ldots,F_m; k)$ has a solution, then Step 2 will return such a solution.

On the other hand, if there exists an agreement forest $F^*$ for $(F,F_3,\ldots,F_m;k)$ with Ord$(F^*) \leq k$, then, obviously, $F^*$ is also a solution for $(F_1,F_2,F_3,\ldots,F_m;k)$. Thus, every solution for $(F,F_3,\ldots,F_m;k)$ is also a solution for $(F_1,F_2,F_3,\ldots,F_m;k)$. That is, if $(F_1,F_2,F_3,\ldots,F_m; k)$ has no solution, Step 2 could not return a solution.

The theorem is proved.
\end{proof}
\end{theorem}

In the following two subsections, we will discuss the detail ways of how to construct all maximal-AFs for two rooted general $X$-forests and for two unrooted general $X$-forests separately. Then, combining the Main Algorithm, we can give the detail parameterized algorithms for {\sc para-rMaf} and {\sc para-uMaf}.

\subsection{Parameterized Algorithm for {\sc para-rMaf}}

Two leaves of a rooted general $X$-forest are siblings if they have a common parent. A {\it sibling set} is set of leaves that are siblings. A {\it maximal sibling set} (abbr. MSS) $S$ is a sibling set that the common parent $p$ of $S$ has degree either $|S|$ if $p$ has no parent or $|S|+1$ if $p$ has a parent.

In this subsection, we present the way of enumerating all maximal-AFs for two rooted general $X$-forests $F_1$ and $F_2$. Let $F^*$ be a fixed maximal-AF for $F_1$ and $F_2$. We begin with a simple lemma.

\begin{lemma}
\label{simpleroot}
Let $F_1$ and $F_2$ be two rooted general $X$-forests. If $F_2$ has no MSS, then $F_1$ and $F_2$ have the unique maximal-AF which can be constructed in linear time.

\begin{proof}
If $F_2$ has no MSS, then $F_2$ has at most one edge. If $F_2$ has no edge, then all connected components of $F_2$ are single-vertex trees and $F_2$ itself is the unique maximal-AF for $F_1$ and $F_2$.

If $F_2$ has one edge, then all connected components of $F_2$ are single-vertex trees except one that is a single-edge tree whose root is $\rho$ with a unique child that is labeled by a label $\tau$. If $\rho$ and $\tau$ are in the same connected component in $F_1$, then the unique maximal-AF for $F_1$ and $F_2$ is $F_2$ itself; otherwise, the unique maximal-AF for $F_1$ and $F_2$ consists of only single-vertex trees, each is labeled by an element in $X$.
\end{proof}
\end{lemma}

By Lemma~\ref{simpleroot}, in the following discussion, we will assume that $F_2$ has an MSS $S$. Because we assumed that all instances in our discussion are strongly reducible, so none of labels in $S$ is a single-vertex tree in $F_1$; otherwise, Reduction Rule 1 can remove the edge incident to the label in $F_2$ which is a single-vertex tree in $F_1$. In the following, we consider all possible cases for the labels of $S$ in $F_1$. Since $|S| \geq 2$, we can assume that labels $a$ and $b$ belong to $S$.

\smallskip

\noindent{\bf Case 1.} All labels in $S$ consist an MSS in $F_1$.

In this case, $F_1$ and $F_2$ have the same local structure in term of $S$, which consists of the labels in $S$ and the parent of $S$. Thus, in the further processing of $F_1$ and $F_2$, the local structure remains unchanged. Therefore, we can treat it as an un-decomposable structure. Note that $F^*$ also have the local structure.

\smallskip

\noindent{\bf Step 1.} Group all labels in $S$ and their parent into an un-decomposable structure, and mark the unit with the same label in $F_1$ and $F_2$.

\smallskip

To implement Step 1, we simply remove all labels in $S$ and label the parent with $\underline{S}$, where $\underline{S}$ is a combination of the labels in $S$ (e.g., assume $S=\{a,b,c\}$, then $\underline{S}=\underline{abc}$). In the further processing of $F_1$ and $F_2$, we can treat $\underline{S}$ as a new leaf in the forests. This step not only changes the structures of $F_1$ and $F_2$, but also replaces the label-set $X$ with a new label-set $(X \setminus S) \cup \{\underline{S}\}$. If we also apply this operation on the maximal-AF $F^*$, then the new $F^*$ remains a maximal-AF for the new $F_1$ and $F_2$.

\smallskip

\noindent{\bf Case 2.} All labels in $S$ are siblings in $F_1$.

\smallskip

Let $p_1$ be the common parent of $S$ in $F_1$. And let $V=\{v_1,\ldots,v_r\}$ be the set that contains all vertices whose parent is $p_1$ in $F_1$, except the labels in $S$. Set $V$ could not be an empty set, otherwise, the labels in $S$ would consist an MSS in $F_1$, which satisfies the condition of Case 1. There are three situations for $a$ and $b$ in $F^*$.

Situation 1. $a$ is a single-vertex tree in $F^*$. Thus, removing the edge incident to $a$ in $F_1$ and $F_2$ keeps $F^*$ still a maximal-AF for $F_1$ and $F_2$.

Situation 2. $b$ is a single-vertex tree in $F^*$. Thus, removing the edge incident to $b$ in $F_1$ and $F_2$ keeps $F^*$ still a maximal-AF for $F_1$ and $F_2$.

Situation 3. Neither $a$ nor $b$ is a single-vertex tree in $F^*$. Because $a$ and $b$ are siblings in $F_2$, so $a$ and $b$ are siblings in $F^*$. Moreover, for this situation, we have the following lemma.

\begin{lemma}
\label{allsiblings-rooted}
Let $F_1$ and $F_2$ be two rooted general $X$-forests, and let $S$ be an MSS of $F_2$ that all labels in $S$ are siblings in $F_1$. For any maximal-AF $F$ for $F_1$ and $F_2$, if there are two labels in $S$ that are siblings in $F$, then all labels in $S$ consist an MSS in $F$.

\begin{proof}
Suppose that labels $a$ and $b$ belong to $S$ and $a$ and $b$ are siblings in $F$. At first, we show that all labels in $S$ are siblings in $F$. There are two cases based on the cardinality of $S$.

Case (i): $|S|=2$. Then, $S=\{a,b\}$. Obviously, this case holds true.

Case (ii): $|S| \geq 3$. Assume that label $c \in S$. If $a$ and $c$ are in different connected components in $F$, then because $a$ and $c$ are siblings in $F_2$, so at least one of $e_a$ and $e_c$ can not be in $F$, where $e_a$ and $e_c$ are the edges that incident to $a$ and $c$ in $F_2$, respectively. Therefore, at least one of $a$ and $c$ is a single-vertex in $F$. But $a$ and $b$ are siblings in $F$, so $a$ is not a single-vertex tree in $F$, thus, $c$ is a single-vertex tree in $F$. By attaching the single-vertex tree $c$ to the common parent of $a$ and $b$ in $F$, we could get an agreement forest for $F_1$ and $F_2$ that consists of fewer trees, which contracts the fact that $F$ is a maximal-AF for $F_1$ and $F_2$. Thus, $a$ and $c$ must be in the same connected component in $F$.
Then, because $a$ and $c$ are siblings in $F_2$, so $a$ and $c$ are also siblings in $F$. Therefore, all labels in $S$ are siblings in $F$.

Now we show that the labels in $S$ consist an MSS in $F$. Since $F$ is a subforest of $F_2$, the parent of $S$ in $F_2$ corresponds to the the parent of $S$ in $F$. And because the parent of $S$ in $F_2$ has $|S|$ children, so the parent of $S$ in $F$ has at most $|S|$ children. Therefore, the labels in $S$ consist an MSS in $F$.
\end{proof}
\end{lemma}

Let $E_V$ be the set that contains all edges $[p_1,v_i]$, $1 \leq i \leq r$. By Lemma~\ref{allsiblings-rooted}, all edges in $E_V$ could not be in $F^*$. Therefore, in Situation 3, removing the edges in $E_V$ from $F_1$ keeps $F^*$ still a maximal-AF for $F_1$ and $F_2$. Summarizing above analysis, we apply the following step. One of these following three branches keeps $F^*$ a maximal-AF for the new $F_1$ and $F_2$.

\smallskip

\noindent{\bf Step 2.}
(branch-1) remove the edge incident to $a$ in both $F_1$ and $F_2$;\\
\hspace*{15mm}(branch-2) remove the edge incident to $b$ in both $F_1$ and $F_2$;\\
\hspace*{15mm}(branch-3) remove the edges in $E_V$.

\smallskip

\noindent{\bf Case 3.} Some labels in $S$ are not siblings in $F_1$.

\smallskip

W.l.o.g., we assume that $a$ and $b$ are not siblings in $F_1$. Let $p_2$ be the common parent of $a$ and $b$ in $F_2$.

\smallskip

\noindent{\bf Subcase 3.1.} $a$ and $b$ are not in the same connected component in $F_1$.

\smallskip

Because $a$ and $b$ are not in the same connected component in $F_1$, so $a$ and $b$ cannot be in the same connected component in $F^*$. Thus, at least one of edges $[a,p_2]$ and $[b,p_2]$ in $F_2$ could not be in $F^*$. Therefore, at least one of $a$ and $b$ is a single-vertex tree.

\smallskip

\noindent{\bf Step 3.1.}
(branch-1) remove the edge incident to $a$ in both $F_1$ and $F_2$;\\
\hspace*{19mm}(branch-2) remove the edge incident to $b$ in both $F_1$ and $F_2$.

\smallskip

One of the two branches must keep $F^*$ still a maximal-AF for the new $X$-forests $F_1$ and $F_2$.

\smallskip

\noindent{\bf Subcase 3.2.} $a$ and $b$ are in the same connected component in $F_1$.

\smallskip

Let $P = \{a,c_1,\ldots,c_t,b\}$ be the path in $F_1$ that connects $a$ and $b$, in which $c_h$ is the least common ancestor of $a$ and $b$, $1 \leq h \leq t$. And let $E_p$ be the edge set that contains all edges in $F_1$ that incident to $c_i$, $1 \leq i \leq t, i \neq h$, but not on the path $P$. There are also three situations for $a$ and $b$ in $F^*$, which are the same as that for Case 2. The first two situations that either $a$ or $b$ is a single-vertex tree in $F^*$ again cause removing the edge incident to $a$ or $b$ in $F_1$ and $F_2$.

For situation 3: neither $a$ nor $b$ is a single-vertex tree in $F^*$, we have to analyze detailly here. In this situation, again $a$ and $b$ are siblings in $F^*$. Moreover, all edges in $E_p$ could not be in $F^*$. Note that because the subtrees in an $X$-forest preserve the ancestor-descendant relation, the edges incident to $c_h$, but not on $P$ could not be removed in this subcase (there may be more than one such edge).

\smallskip

\noindent{\bf Step 3.2.}(branch-1) remove the edge incident to $a$ in both $F_1$ and $F_2$;\\
\hspace*{17mm}(branch-2) remove the edge incident to $b$ in both $F_1$ and $F_2$;\\
\hspace*{17mm}(branch-3) remove the edges in $E_p$.

\smallskip

One of these three branches keeps $F^*$ a maximal-AF for the new $F_1$ and $F_2$.

\smallskip

For two given rooted general $X$-forests $F_1$ and $F_2$, if we iteratively apply the above process, branching accordingly based on the cases, then the process will end up with a pair $(F_1, F_2)$ in which $F_2$ contains no MSS. When this occurs, the process applies the following step:

\smallskip

\noindent {\bf Final Step.} If $F_2$ contains no MSS, then construct the (unique) maximal-AF $F^*$ for $F_1$ and $F_2$, and convert $F^*$ into an agreement forest for the original $F_1$ and $F_2$.

\smallskip

When $F_2$ contains no MSS, by Lemma~\ref{simpleroot}, we can construct the unique maximal-AF $F^*$ for $F_1$ and $F_2$ in linear time. The forest $F^*$ may not be a subforest of the original $F_1$ and $F_2$ because Step 1 shrinks labels. For this, we should ``expand'' the shrunk labels, in a straightforward way. Note that this expanding process may be applied iteratively, but in linear time.

Summarizing the above discussion, we conclude with the following lemma.

\begin{lemma}
\label{rootlem}
Let $F_1$ and $F_2$ be two rooted general $X$-forests. If we apply Steps 1-3.2 iteratively until $F_2$ contains no MSS, then for every maximal-AF $F^*$ for the original $F_1$ and $F_2$, at least one of the branches in the process produces the maximal-AF $F^*$ in its Final Step.

\begin{proof}
Fix a maximal-AF $F^*$ for the original $F_1$ and $F_2$. By the above analysis, for each of the cases, at least one of the branches in the corresponding step keeps $F^*$ a maximal-AF for $F_1$ and $F_2$. Moreover, when $F_2$ contains no MSS, the maximal-AF for $F_1$ and $F_2$ becomes unique. Combining these two facts, we can conclude that at least one of the branches in the process ends up with a pair $F_1$ and $F_2$ whose maximal-AF, after the final step, is $F^*$. Since $F^*$ is an arbitrary maximal-AF for $F_1$ and $F_2$, the lemma is proved.
\end{proof}
\end{lemma}

Now, according to the discussion given above and Main-Algorithm, we can present the detail parameterized algorithm for the {\sc para-rMaf} problem, which is presented in Figure~\ref{algrooted-para}.

\begin{figure}[gpath3]
\setbox4=\vbox{\hsize28pc \noindent\strut \begin{quote}
\vspace*{-6mm} \footnotesize

{\bf Algorithm} Alg-{\sc para-rMaf}$(F_1, F_2, \ldots, F_m; k)$\\
Input: a collection $\{F_1, F_2, \ldots, F_m\}$ of rooted general $X$-forests, $m \geq 1$,\\
   \hspace*{10mm} and a parameter $k$\\
Output: an agreement forest $F^*$ for $\{F_1, F_2, \ldots, F_m\}$ with \\
    \hspace*{13mm} Ord$(F^*) \leq k$ if such an $F^*$ exists

1. {\bf if} $(m=1)$ {\bf then} {\bf if} (Ord$(F_1) \leq k$) {\bf then} return $F_1$
      {\bf else} return(`no');\\
2. {\bf if} (Ord$(F_1) > k$) {\bf then} return(`no');\\
3. {\bf apply} Reduction Rule 1 on $F_1$ and $F_2$ if possible;\\
4. {\bf if} $F_2$ has no MSS {\bf then} let $F'$ be the maximal-AF for\\
   \hspace*{3mm} $F_1$ and $F_2$; return Alg-{\sc para-rMaf}$(F', F_3, \ldots, F_m; k)$;\\
5. let $S$ be an MSS in $F_2$;  /** assume labels $a$ and $b$ belong to $S$\\
6. {\bf if} all labels in $S$ consist an MSS in $F_1$\\
\hspace*{3mm} {\bf then} group all labels in $S$ into a new leaf $\underline{S}$ in both $F_1$ and $F_2$;\\ \hspace*{4mm}return Alg-{\sc para-rMaf}$(F_1, F_2, \ldots, F_m; k)$;\\
7. {\bf if} all labels in $S$ are siblings in $F_1$, let $V=\{v_1,\ldots,v_r\}$ be the set that\\
\hspace*{3mm} contains all vertices which have a common parent with $a$ in $F_1$,\\
\hspace*{3mm} except the labels in $S$, $r \geq 1$, {\bf then} branch:\\
\hspace*{14mm} 1.\ make $a$ a single-vertex tree in both $F_1$ and $F_2$;\\
                  \hspace*{18mm} return Alg-{\sc para-rMaf}$(F_1, F_2, \ldots, F_m; k)$;\\
\hspace*{14mm} 2.\ make $b$ a single-vertex tree in both $F_1$ and $F_2$;\\
                  \hspace*{18mm} return Alg-{\sc para-rMaf}$(F_1, F_2, \ldots, F_m; k)$;\\
\hspace*{14mm} 3.\ remove all edges that between $v_i$ and the common parent of $S$ in $F_1$, for $1 \leq i \leq r$;\\
\hspace*{18mm} return Alg-{\sc para-rMaf}$(F_1, F_2, \ldots, F_m; k)$;\\
8. {\bf if} there are two labels in $S$ that are not siblings in $F_1$  /** assume $a$ and $b$ are not siblings\\
8.1. \hspace*{3mm}{\bf if} $a$ and $b$ are in different connected components in $F_1$,\\
\hspace*{8mm} {\bf then} branch:\\
\hspace*{14mm} 1.\ make $a$ a single-vertex tree in both $F_1$ and $F_2$;\\
                  \hspace*{18mm} return Alg-{\sc para-rMaf}$(F_1, F_2, \ldots, F_m; k)$;\\
\hspace*{14mm} 2.\ make $b$ a single-vertex tree in both $F_1$ and $F_2$;\\
                 \hspace*{18mm} return Alg-{\sc para-rMaf}$(F_1, F_2, \ldots, F_m; k)$;\\
8.2. \hspace*{3mm}{\bf if} $a$ and $b$ are in the same connected components in $F_1$,\\
\hspace*{9mm}let $P = \{a, c_1, \ldots, c_r, b\}$ be the unique path in $F_1$ connecting\\
    \hspace*{8mm} $a$ and $b$, $r \geq 2$, {\bf then} branch:\\
\hspace*{14mm} 1.\ make $a$ a single-vertex tree in both $F_1$ and $F_2$;\\
                  \hspace*{18mm} return Alg-{\sc para-rMaf}$(F_1, F_2, \ldots, F_m; k)$;\\
\hspace*{14mm} 2.\ make $b$ a single-vertex tree in both $F_1$ and $F_2$;\\
                  \hspace*{18mm} return Alg-{\sc para-rMaf}$(F_1, F_2, \ldots, F_m; k)$;\\
\hspace*{14mm} 3.\ remove all edges in $F_1$ that are not on $P$ but incident to\\
                  \hspace*{18mm} a vertex in $P$, except the ones incident to the least common\\
                 \hspace*{18mm} ancestor of $a$ and $b$; return Alg-{\sc para-rMaf}$(F_1, F_2, \ldots, F_m; k)$.
\end{quote}
\vspace*{-6mm} \strut} $$\boxit{\box4}$$ \vspace*{-9mm}
\caption{Algorithm for the {\sc para-rMaf} problem}
\label{algrooted-para}
\end{figure}

We consider the correctness and the complexity of the algorithm. To make our discussion more specific, we denote by $(\bar{F}_1, \bar{F}_2, \ldots, \bar{F}_m; k)$ the original input to the algorithm, and initiate with $F_i = \bar{F}_i$ for
$1 \leq i \leq m$.

The algorithm is a branch-and-search process, in which Step 7, Step 8.1, and Step 8.2 contain branches. The execution of the algorithm can be depicted by a search tree $\cal T$ whose leaves correspond to conclusions or solutions generated by the algorithm based on different branches. Each internal node of the search tree $\cal T$ corresponds to a branch in the search process at Step 7, or Step 8.1, or Step 8.2 based on an instance of the problem. The root of the tree $\cal T$ is on the instance that is the original input to the algorithm. We will call a path from the root to a leaf in the search tree $\cal T$ a {\it computational path} in the process, which corresponds to a particular sequence of executions in the algorithm that leads to a conclusion or solution. The algorithm returns an agreement forest for the original input if and only if there is a computational path that outputs the forest.

We first study the correctness of the algorithm.

According to Step 1, the algorithm is correct when $m = 1$. Therefore, we will assume that $m > 1$ and that the algorithm is correct when the input contains no more than $m-1$ $X$-forests.

If Ord$(F_1) > k$, then an MAF $F'$ for $(F_1, F_2, \ldots, F_m)$, which is a subforest of $F_1$, must have Ord$(F') > k$. Thus, the instance $(F_1, F_2, \ldots, F_m; k)$ is a `no'. Step 2 is correct.
By Lemma~\ref{Reduction Rule 1}, Step 3 is also correct.

If $F_2$ has no MSS, then by Lemma~\ref{simpleroot}, the unique maximal-AF $F'$ for $F_1$ and $F_2$ can be constructed in linear time. Since $F'$ is the unique maximal-AF for $F_1$ and $F_2$, by Lemma~\ref{maximal-MAF}, the instances $(F_1, F_2, \ldots, F_m; k)$ and $(F', \ldots, F_m; k)$ have the same set of MAFs. By our induction, the algorithm works correctly on $(F', \ldots, F_m; k)$. Thus, Step 4 is correct.

If the instance $(F_1, F_2, \ldots, F_m; k)$ reaches step 5, then the $X$-forest $F_2$ has an MSS $S$, and none of labels in $S$ is a single-vertex tree in $F_1$. Steps 6-8 are applied on the $X$-forests $F_1$ and $F_2$ recursively (during the recursion, Step 3 may also be applied). These steps remove edges in $F_1$ and $F_2$ thus reduce the sizes of the forests (Step 6 does not remove edges, but it reduces the size of $F_1$ and $F_2$ without changing their essential structures). Thus, the steps keep the situation, recursively, that $F_1$ and $F_2$ are subforests of $\bar{F}_1$ and $\bar{F}_2$, respectively. This means that during the process of
these steps, every agreement forest for $\{F_1, F_2, \ldots, F_m\}$ remains an agreement forest for the original $\{\bar{F}_1, \bar{F}_2, \ldots, \bar{F}_m\}$. These steps continue until the condition in either Step 2 or Step 4 is met. By the discussion above, Step 2 or Step 4 then will return a correct solution to the instance $(F_1, F_2, \ldots, F_m; k)$, which is either an answer `no', or an agreement forest $F^*$ for $\{F_1, F_2, \ldots, F_m\}$ with Ord$(F^*) \leq k$, which is also a solution to the original input $(\bar{F}_1, \bar{F}_2, \ldots, \bar{F}_m; k)$. Thus, no computational path in the algorithm can output an $X$-forest that is not a solution to the original input $(\bar{F}_1, \bar{F}_2, \ldots, \bar{F}_m; k)$. In particular, if the original input $(\bar{F}_1, \bar{F}_2, \ldots, \bar{F}_m; k)$ is a `no' for the {\sc para-rMaf} problem, then the algorithm Alg-{\sc para-rMaf} must return an answer `no'. On the other hand, suppose that $(\bar{F}_1, \bar{F}_2, \ldots, \bar{F}_m; k)$ is a `yes' and $\{\bar{F}_1, \bar{F}_2, \ldots, \bar{F}_m\}$ has an MAF $F^*$ with Ord$(F^*) \leq k$. Let $F'$ be the maximal-AF for $\bar{F}_1$ and $\bar{F}_2$ that has $F^*$ as a subforest (note Ord$(F') \leq \mbox{Ord}(F^*) \leq k$). By Lemma~\ref{rootlem}, there is a computational path that starts with $F_1 = \bar{F}_1$ and $F_2 = \bar{F}_2$, and applies Steps 6-8 recursively until $F_1$ and $F_2$ satisfy the condition of Step 4. Step 4 then constructs the maximal-AF $F'$ for $\bar{F}_1$ and $\bar{F}_2$. By Lemma~\ref{maximal-MAF} and our induction, the recursive call in Step 4 will return an agreement forest $F$ for $\{\bar{F}_1, \bar{F}_2, \ldots, \bar{F}_m\}$ with Ord$(F) \leq k$. Therefore, the algorithm also works correctly in this case.

This completes the proof of the correctness of the algorithm. Now we consider the complexity of the algorithm. Because of Step 7, Step 8.1, and Step 8.2, each branch in the search tree $\cal T$ can make at most three ways. Moreover, by examining Steps 7, 8.1, and 8.2, it is easy to verify that between two consecutive branches in a computational path, the value Ord$(F_1)$ is increased by at least $1$. Since the algorithm will stop at Step 2 when Ord$(F_1) > k$, each computational path in the search tree $\cal T$ can go through at most $k$ branches. As a consequence, the number of leaves in the search tree $\cal T$ is bounded by $3^k$. It takes time $O(n)$ to judge whether two labels are in the same connected component, where $n$ is the size of label-set $X$. Thus, it is easy to verify that between two consecutive branches, the computational path takes time $O(n^2m)$, where $m$ is the number of $X$-forests in the original input instance. Summarizing all these together, we conclude that the algorithm Alg-{\sc para-rMaf}$(F_1, F_2, \ldots, F_m; k)$ has its running time bounded by $O(3^k n^2m)$.

\begin{theorem}
The {\sc para-rMaf} problem can be solved in time $O(3^k n^2m)$, where $n$ is the size of label-set $X$ and $m$ is the number of $X$-forests in the input instance.
\end{theorem}

\subsection{Parameterized Algorithm for {\sc para-uMaf}}

The discussion for the {\sc para-uMaf} problem on the instance $(F_1,F_2,\ldots,F_m;k)$ is similar to that for the {\sc para-rMaf} problem. However, since unrooted $X$-forests preserve no ancestor-descendant relation, there is a little difference.

Two leaves in an unrooted $X$-forest are {\it siblings} if either they are connected by an edge or they have a common neighbor. A {\it sibling set} is a set of leaves that are siblings. A sibling set $S$ is {\it maximal} (abbr. MSS) if either $S$ is the label set of a single-edge tree or the common neighbor of $S$ has degree at most $|S|+1$.

An unrooted $X$-forest with no MSS has an even simpler structure: all its connected components are single-vertex trees. Thus, we have the following lemma, which is similar to Lemma~\ref{simpleroot}.

\begin{lemma}
\label{simpleunroot}
Let $F_1$ and $F_2$ be two unrooted general $X$-forests. If $F_2$ has no MSS, then $F_1$ and $F_2$ have the unique maximal-AF which can be constructed in linear time.
\end{lemma}

Thus, in the following discussion, we will assume that $F_2$ has an MSS $S$ that none of labels in $S$ is a single-vertex tree in $F_1$. And, we will assume that labels $a$ and $b$ belong to $S$. In the following, we consider all possible cases for labels of $S$ in $F_1$. Let $F^*$ be a fixed maximal-AF for $F_1$ and $F_2$.

\smallskip

\noindent{\bf Case 1.} All labels in $S$ consist an MSS in $F_1$.

\smallskip

In this case, we also treat $S$ as an un-decomposable structure.

\smallskip

\noindent{\bf Step 1.} Group all labels in $S$ (and their common neighbor if $S$ is not the label set of a single-edge tree) into an un-decomposable structure, and mark the unit with the same label in $F_1$ and $F_2$.

\smallskip

To implement Step 1, if $S$ is the label set of a single-edge tree, then combine the labels into a single vertex that labeled by $\underline{S}$; otherwise, simply remove all labels in $S$ and label the common neighbor of $S$ with $\underline{S}$. This step not only changes the structures of $F_1$ and $F_2$, but also replaces the label-set $X$ with a new label-set $(X \setminus S) \cup \{\underline{S}\}$. If we also apply this operation in the maximal-AF $F^*$, then the new $F^*$ remains a maximal-AF for the new $F_1$ and $F_2$.

\smallskip

\noindent{\bf Case 2.} All labels in $S$ are siblings in $F_1$.

\smallskip

Obviously, the common neighbor $p$ of $S$ in $F_1$ has degree not less than $|S|+2$. Let $V = \{v_1,\ldots,v_r\}$ be the vertex set that contains all vertices that are neighbors of $p$, except the labels in $S$. Obviously, $r \geq 2$. Let $e_1$ be the edge that between $v_1$ and $p$ in $F_1$, and let $e_r$ be the edge that between $v_r$ and $p$ in $F_1$. There are three situations for $a$ and $b$ in $F^*$.

Situation 1. $a$ is a single-vertex tree in $F^*$. Thus, removing the edge incident to $a$ in $F_1$ and $F_2$ keeps $F^*$ still a maximal-AF for $F_1$ and $F_2$.

Situation 2. $b$ is a single-vertex tree in $F^*$. Thus, removing the edge incident to $b$ in $F_1$ and $F_2$ keeps $F^*$ still a maximal-AF for $F_1$ and $F_2$.

Situation 3. neither $a$ nor $b$ is a single-vertex tree in $F^*$. Because $a$ and $b$ are siblings in $F_2$, so $a$ and $b$ are siblings in $F^*$. Moreover, in this situation, we have the following lemma.

\begin{lemma}
\label{allsiblings-unrooted}
Let $F_1$ and $F_2$ be two unrooted general $X$-forests, and let $S$ be an MSS of $F_2$ that all labels in $S$ are siblings in $F_1$. For any maximal-AF $F$ for $F_1$ and $F_2$, if there are two labels in $S$ that are siblings in $F$, then all labels in $S$ consist an MSS in $F$.

\begin{proof}
Suppose that labels $a$ and $b$ belong to $S$ and $a$ and $b$ are siblings in $F$. At first, we show that all labels in $S$ are siblings in $F$.

Case (i): $|S|=2$. Then, $S=\{a,b\}$. Obviously, this case holds true.

Case (ii): $|S| \geq 3$. Suppose that label $c \in S$. If $a$ and $c$ are in different connected components in $F$, then because $a$ and $c$ are siblings in $F_2$, so at least one of $e_a$ and $e_c$ cannot be in $F$, where $e_a$ and $e_c$ are the edges that incident to $a$ and $c$ in $F_2$, respectively. Therefore, at least one of $a$ and $c$ is a single-vertex tree in $F$. Since $a$ and $b$ are in the same connected component in $F$, $a$ is not a single-vertex tree in $F$, thus, $c$ is a single-vertex tree in $F$. By attaching the single-vertex tree $c$ to the common neighbor of $a$ and $b$ in $F$ (if $a$ and $b$ are two labels of a single-edge tree in $F$, then subdividing the edge between $a$ and $b$ by a new vertex and attaching $c$ to the new vertex), we could get an agreement forest for $F_1$ and $F_2$ that consists of fewer trees, which contracts the fact that $F$ is a maximal-AF for $F_1$ and $F_2$. Thus, $a$ and $c$ must be in the same connected component in $F$. Then, because $a$ and $c$ are siblings in $F_2$, so $a$ and $c$ are siblings in $F$. Therefore, all labels in $S$ are siblings in $F$.

Now we show that the labels in $S$ consist an MSS. The common neighbor of $S$ in $F_2$ corresponds to the common neighbor of $S$ in $F$. Because the common neighbor of $S$ in $F_2$ has degree at most $|S|+1$, so the common neighbor of $S$ in $F$ also has degree at most $|S|+1$. Thus, the labels in $S$ consist an MSS in $F$.
\end{proof}
\end{lemma}

By Lemma~\ref{allsiblings-unrooted}, in Situation 3, all labels in $S$ are siblings in $F^*$. If the common neighbor of $S$ in $F_2$ has degree $|S|$, then the common neighbor of $S$ in $F^*$ has degree $|S|$, edges $e_1$ and $e_r$ in $F_1$ cannot be in $F^*$, both of them should be removed. If the common neighbor of $S$ in $F_2$ has degree $|S|+1$, then the common neighbor of $S$ in $F^*$ has degree at most $|S|+1$. But the common neighbor of $S$ in $F_1$ has degree at least $|S|+2$, thus, at least one of $e_1$ and $e_r$ in $F_1$ could not be in $F^*$. However, the subtrees in unrooted $X$-forests do not preserve any ancestor-descendant relation, we cannot decide which one of $e_1$ and $e_r$ in $F_1$ should be removed. Therefore, we can branch by removing $e_1$ or $e_r$.

In Situation 3, whether or not the common neighbor of $S$ in $F_2$ has degree $|S|$, branching by removing $e_1$ or $e_r$ always is right.

Summarizing the above analysis, we can apply the following step. One of these following branches keeps $F^*$ a maximal-AF for the new $F_1$ and $F_2$.

\smallskip

\noindent{\bf Step 2.}
(branch-1) remove the edge incident to $a$ in both $F_1$ and $F_2$;\\
\hspace*{15mm}(branch-2) remove the edge incident to $b$ in both $F_1$ and $F_2$;\\
\hspace*{15mm}(branch-3) remove the edge $e_1$ in $F_1$;\\
\hspace*{15mm}(branch-4) remove the edge $e_r$ in $F_1$.

\smallskip

\noindent{\bf Case 3.} Some labels in $X$ are not siblings in $F_1$.

\smallskip

W.l.o.g., we assume $a$ and $b$ are not siblings in $F_1$.

\smallskip

\noindent{\bf Subcase 3.1.} $a$ and $b$ are not in the same connected component in $F_1$.

\smallskip

Again that at least one of $a$ and $b$ must be a single-vertex tree in $F^*$. We can apply the following step.

\smallskip

\noindent{\bf Step 3.1.}
(branch-1) remove the edge incident to $a$ in both $F_1$ and $F_2$;\\
\hspace*{19mm}(branch-2) remove the edge incident to $b$ in both $F_1$ and $F_2$.

\smallskip

\noindent{\bf Subcase 3.2.} $a$ and $b$ are in the same connected component in $F_1$.

\smallskip

Let $P=\{a,c_1,c_2,\ldots,c_r,b\}$ be the unique path that connects $a$ an $b$ in $F_1$, $r \geq 2$. There are also three situations for $a$ and $b$ in $F^*$, which are the same as that for Case 2. The first two situations that either $a$ or $b$ is a single-vertex tree in $F^*$ again cause removing the edge incident to $a$ or $b$ in both $F_1$ and $F_2$.

For Situation 3: neither $a$ nor $b$ is a single-vertex tree in $F^*$, we have to analyze in detail. Because $a$ and $b$ are siblings in $F_2$, so $a$ and $b$ are siblings in $F^*$.

If $a$ and $b$ are connected by an edge in $F^*$, then all internal vertices in $P$ should be removed by the forced contraction. That is, all the edges that not on the path $P$ but incident to a internal vertex in $P$ cannot be in $F^*$, thus, all these edges should be removed. If $a$ and $b$ have a common neighbor in $F^*$, then only one internal vertex in $P$ can be kept, and all the other internal vertices in $P$ should be removed by the forced contraction. Since the subtrees in unrooted $X$-forests do not preserve any ancestor-descendant relation, we do not know which one of the internal vertices in $P$ should be kept. On the other side, we know that at least one of $c_1$ and $c_r$ should be removed by the forced contraction. That is, either the edges that incident to $c_1$ but not on $P$ or the edges that incident to $c_r$ but not on $P$ should be removed. Therefore, we can branch by removing the edges that incident to $c_1$ but not on $P$ or the edges that incident to $c_r$ but not on $P$. In this situation, whether or not $a$ and $b$ are connected by an edge in $F^*$, this branching way is always right.

Summarizing the above analysis, we can apply the following step. One of these branches keeps $F^*$ a maximal-AF for the new $F_1$ and $F_2$.

\smallskip

\noindent{\bf Step 3.2.}
(branch-1) remove the edge incident to $a$ in both $F_1$ and $F_2$;\\
\hspace*{19mm}(branch-2) remove the edge incident to $b$ in both $F_1$ and $F_2$;\\
\hspace*{19mm}(branch-3) the edges that incident to $c_1$ but not on $P$ in $F_1$;\\
\hspace*{19mm}(branch-4) the edges that incident to $c_r$ but not on $P$ in $F_1$.

\smallskip

For two given unrooted general $X$-forests $F_1$ and $F_2$, if we iteratively apply the above process, branching accordingly based on the cases, then the process will end up with a pair $(F_1, F_2)$ in which $F_2$ contains no MSS. When this occurs, we again apply the following step:

\smallskip

\noindent {\bf Final Step.} If $F_2$ contains no MSS, then construct the (unique) maximal-AF $F^*$ for $F_1$ and $F_2$, and convert $F^*$ into an agreement forest for the original $F_1$ and $F_2$.

\smallskip

The above analysis finally gives the following conclusion, whose proof is exactly the same as that of Lemma~\ref{rootlem} in Subsection 4.1.

\begin{lemma}
\label{unrootlem}
Let $F_1$ and $F_2$ be two unrooted general $X$-forests. If we apply Steps 1-3.2 iteratively until $F_2$ contains no MSS, then for every maximal-AF $F^*$ for the original $F_1$ and $F_2$, at least one of the branches in the process produces the maximal-AF $F^*$ in its Final Step.
\end{lemma}

\begin{figure}[h]
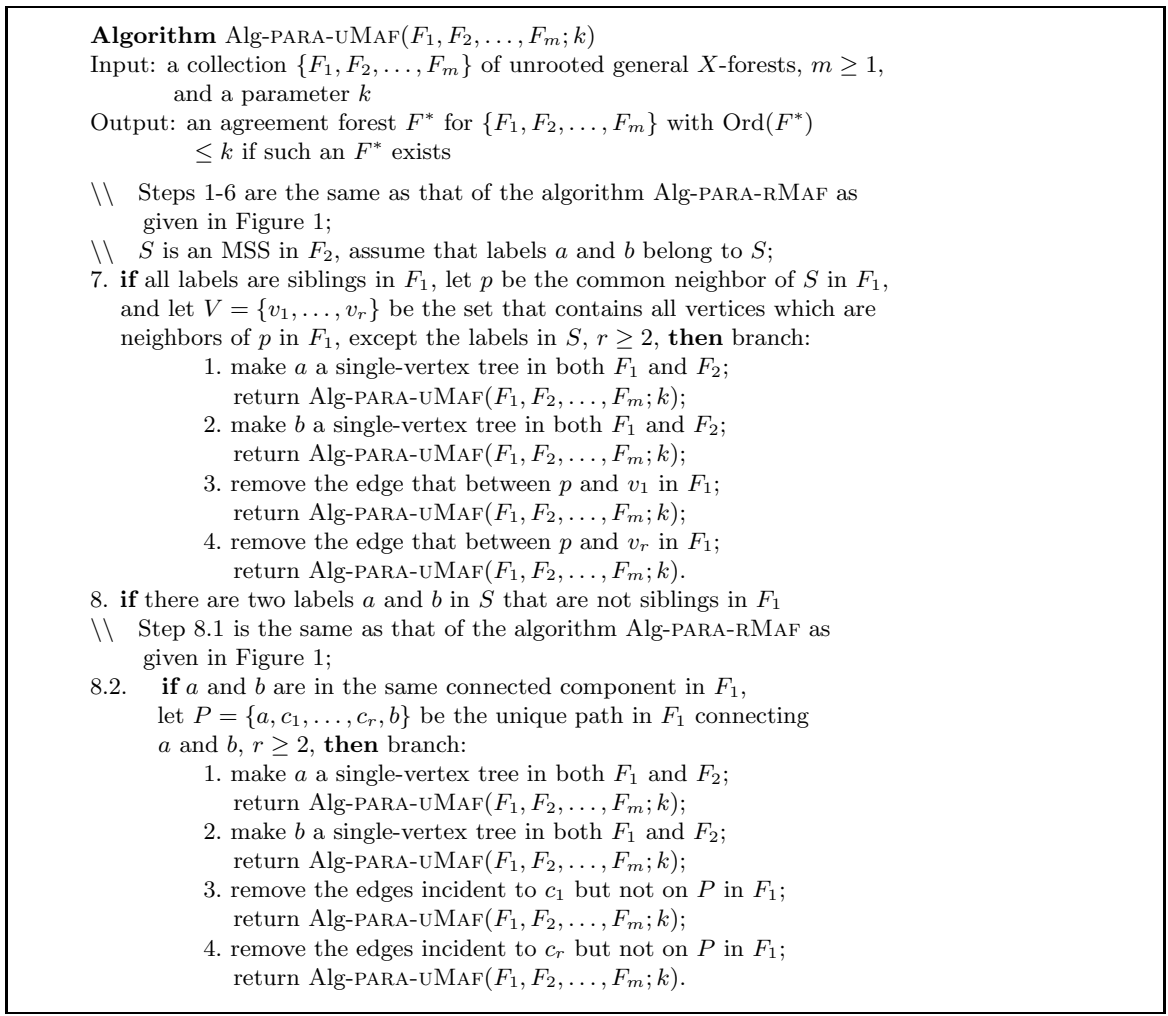

\setbox4=\vbox{\hsize28pc \noindent\strut \begin{quote}
\vspace*{-6mm} \footnotesize

{\bf Algorithm} Alg-{\sc para-uMaf}$(F_1, F_2, \ldots, F_m; k)$\\
Input: a collection $\{F_1, F_2, \ldots, F_m\}$ of unrooted general $X$-forests, $m \geq 1$,\\
       \hspace*{10mm} and a parameter $k$\\
Output: an agreement forest $F^*$ for $\{F_1, F_2, \ldots, F_m\}$ with Ord$(F^*)\\
        \hspace*{13mm} \leq k$ if such an $F^*$ exists

$\backslash\backslash$ \hspace*{1mm} Steps 1-6 are the same as that of the algorithm Alg-{\sc para-rMaf} as\\
        \hspace*{6mm} given in Figure~\ref{algrooted-para}; \\
$\backslash\backslash$ \hspace*{1mm} $S$ is an MSS in $F_2$, assume that labels $a$ and $b$ belong to $S$;\\
7. {\bf if} all labels are siblings in $F_1$, let $p$ be the common neighbor of $S$ in $F_1$,\\
\hspace*{3mm} and let $V=\{v_1,\ldots,v_r\}$ be the set that contains all vertices which are\\
\hspace*{3mm} neighbors of $p$ in $F_1$, except the labels in $S$, $r \geq 2$, {\bf then} branch:\\
\hspace*{14mm} 1.\ make $a$ a single-vertex tree in both $F_1$ and $F_2$;\\
                  \hspace*{18mm} return Alg-{\sc para-uMaf}$(F_1, F_2, \ldots, F_m; k)$;\\
\hspace*{14mm} 2.\ make $b$ a single-vertex tree in both $F_1$ and $F_2$;\\
                  \hspace*{18mm} return Alg-{\sc para-uMaf}$(F_1, F_2, \ldots, F_m; k)$;\\
\hspace*{14mm} 3.\ remove the edge that between $p$ and $v_1$ in $F_1$;\\
\hspace*{18mm} return Alg-{\sc para-uMaf}$(F_1, F_2, \ldots, F_m; k)$;\\
\hspace*{14mm} 4.\ remove the edge that between $p$ and $v_r$ in $F_1$;\\
\hspace*{18mm} return Alg-{\sc para-uMaf}$(F_1, F_2, \ldots, F_m; k)$.\\
8. {\bf if} there are two labels $a$ and $b$ in $S$ that are not siblings in $F_1$\\
$\backslash\backslash$ \hspace*{1mm} Step 8.1 is the same as that of the algorithm Alg-{\sc para-rMaf} as\\
        \hspace*{6mm} given in Figure~\ref{algrooted-para}; \\
8.2. \hspace*{3mm}{\bf if} $a$ and $b$ are in the same connected component in $F_1$,\\
\hspace*{9mm}let $P = \{a, c_1, \ldots, c_r, b\}$ be the unique path in $F_1$ connecting\\
    \hspace*{8mm} $a$ and $b$, $r \geq 2$, {\bf then} branch:\\
\hspace*{14mm} 1.\ make $a$ a single-vertex tree in both $F_1$ and $F_2$;\\
                  \hspace*{18mm} return Alg-{\sc para-uMaf}$(F_1, F_2, \ldots, F_m; k)$;\\
\hspace*{14mm} 2.\ make $b$ a single-vertex tree in both $F_1$ and $F_2$;\\
                  \hspace*{18mm} return Alg-{\sc para-uMaf}$(F_1, F_2, \ldots, F_m; k)$;\\
\hspace*{14mm} 3.\ remove the edges incident to $c_1$ but not on $P$ in $F_1$;\\
                 \hspace*{18mm} return Alg-{\sc para-uMaf}$(F_1, F_2, \ldots, F_m; k)$;\\
\hspace*{14mm} 4.\ remove the edges incident to $c_r$ but not on $P$ in $F_1$;\\
                 \hspace*{18mm} return Alg-{\sc para-uMaf}$(F_1, F_2, \ldots, F_m; k)$.
\end{quote}
\vspace*{-6mm} \strut} $$\boxit{\box4}$$ \vspace*{-9mm}
\caption{Algorithm for the {\sc para-uMaf} problem}
\label{algunrt-para}
\end{figure}

Now we are ready for giving the detail parameterized algorithm for the {\sc para-uMaf} problem, which is presented in Figure~\ref{algunrt-para}.

Similar to the one for {\sc para-rMaf}, the algorithm for the {\sc para-uMaf} problem is a combination of the analysis given in Section 4.2 and the Main-Algorithm. Comparing the analysis for {\sc para-rMaf} given in Section 4.1 and the analysis for {\sc para-uMaf} given in Section 4.2, we can see that they only differ
for Case 2 and Case 3.2: Case 2 and Case 3.2 in Section 4.1 branches into three ways while Case 2 and Case 3.2 in Section 4.2 branch into four ways. Therefore, the two algorithms only need to differ in Step 7 and Step 8.2.

The proof of the correctness for the algorithm Alg-{\sc para-uMaf} proceeds in exactly the same way, based on the analysis in Section 4.2, as that for the algorithm Alg-{\sc para-rMaf}, which was based on the analysis in Section 4.1. For the computational complexity, since Step 7 and Step 8.2 of the algorithm Alg-{\sc para-uMaf} branches into four ways, the search tree $\cal T$ for the algorithm Alg-{\sc para-uMaf} has four-way branches. Therefore, we can conclude that the number of leaves in the search tree $\cal T$ is bounded by $4^k$. All other analysis is the same as that for the algorithm Alg-{\sc para-rMaf}. As a result, we conclude that the algorithm Alg-{\sc para-uMaf} runs in time $O(4^k n^2m)$.

\begin{theorem}
The {\sc para-uMaf} problem can be solved in time $O(4^k n^2m)$, where $n$ is size of the label-set $X$ and $m$ is the number of $X$-forests in the input instance.
\end{theorem}

\section{Approximation Algorithms}

In this section, we will present the approximation algorithms for {\sc app-rMaf} and {\sc app-uMaf} seperately. First of all, we give several related definitions, which apply for both {\sc app-rMaf} and {\sc app-uMaf}.

Let $F$ be an $X$-forest (either rooted or unrooted), and let $E$ be a subset of edges in $F$. Because of forced contraction, we have that Ord$(F \setminus E) \leq \mbox{Ord}(F)+|E|$. An edge-subset $E'$ of $F$ is an {\it essential edge-set} (abbr. ee-set) if Ord$(F \setminus E') =\mbox{Ord}(F)+|E'|$.

Let $E$ be an arbitrary edge-subset of $F$. Define the {\it essential subset} of $E$, denoted by $\widetilde{E}$, to be a subset of $E$ that is an ee-set of $F$ such that $F \setminus E = F \setminus \widetilde{E}$. Obviously, if $E$ itself is an ee-set, then $\widetilde{E} = E$. Note that for an edge-subset $E$ of $F$ which is not an ee-set, there maybe more than one essential subset of $E$.

Our approximation algorithm for {\sc app-Maf} ({\sc app-rMaf} or {\sc app-uMaf}) consist of a sequence of ``meta-steps". An {\it edge-removal meta-step} (or simply meta-step) of an algorithm is a collection of consecutive computational steps in the algorithm that on an instance $(F_1, F_2,\ldots,F_m)$ of {\sc app-Maf} removes certain edges in the forests in $(F_1, F_2,\ldots,F_m)$ (and applies the forced contraction).

Our approximation algorithms for {\sc app-Maf} ({\sc app-rMaf} or {\sc app-uMaf}) have the following general framework.

\begin{figure}[h]
\setbox4=\vbox{\hsize28pc \noindent\strut \begin{quote}
\vspace*{-6mm} \footnotesize

{\bf Algorithm} Alg-{\sc app-Maf}$(F_1, F_2, \ldots, F_m; k)$\\
Input: a collection $\{F_1, F_2, \ldots, F_m\}$ of rooted (unrooted) $X$-forests, $m \geq 1$\\
Output: an agreement forest $F^*$ for $\{F_1, F_2, \ldots, F_m\}$\\
1. {\bf if} $m=1$ {\bf then} return $F_1$;\\
2. {\bf for} $i = 2$ to $m$ {\bf do}\\
3.\hspace*{6mm} {\bf while} $F_1 \neq F_i$\\
\hspace*{14mm} {\bf apply} a meta-step on $F_1$ and $F_i$;\\
4. return $F_1$.
\end{quote}
\vspace*{-6mm} \strut} $$\boxit{\box4}$$ \vspace*{-9mm}
\caption{Algorithm for the {\sc app-Maf}}
\label{alg-app}
\end{figure}

The performance of the approximation algorithm Alg-{\sc app-Maf} heavily depends on the quality of the meta-steps we employ in Step 3 of the algorithm. Thus, we introduce the following concept that measures the quality of a meta-step, where $r \geq 1$ is an arbitrary real number.

\bigskip

\noindent {\bf Definition-R.} Let $I=(F_1,F_2,\ldots,F_m)$ be an instance of {\sc app-Maf} ({\sc app-rMaf} or {\sc app-uMaf}), and let $M$ be an edge-removal meta-step that removes a set $E_M$ of edges in the forests in $I$. Meta-step $M$ keeps a {\it ratio} $r$ if the set $E_M$ contains a subset $E_M^1$ of edges in $F_1$ such that no edge in $E_M \setminus E_M^1$ is in any agreement forest for $(F_1 \setminus E_M^1,F_2,\ldots,F_m)$, and for each agreement forest $F$ for $I$, there always exists an ee-set $E_M^{1,F}$ of $F_1$, $E_M^{1,F} \subseteq E_M^1$, $|E_M^{1,F}| \geq |\widetilde{E_M^1}|/r$, and no edge in $E_M^{1,F}$ is in $F$.

\smallskip

\noindent {\bf Remark 1.} $E_M^1$ contains all edges in $E_M$ that in $F_1$. $E_M^1$ may not be an ee-set of $F_1$, but $E_M^{1,F}$ should be an ee-set of $F_1$.

\smallskip

\noindent {\bf Remark 2.} By definition, if an edge-removal meta-step removes only edges that not in any agreement forest for the instance, then this meta-step keeps ratio $r$ for any $r \geq 1$. Define an edge-removal meta-step is {\it safe} if it keeps ratio $r$ for any $r \geq 1$.

\smallskip

Define the order of an MAF for the instance $(F_1,F_2,\ldots,F_m)$ of {\sc app-Maf} ({\sc app-rMaf} or {\sc app-uMaf}) to be the {\it optimal order} for the instance, denoted $\mbox{Opt}(F_1,F_2,\ldots,F_m)$.

\begin{lemma}
\label{equation}
Let $I=(F_1,F_2,\ldots,F_m)$ be an instance of {\sc app-Maf} ({\sc app-rMaf} or {\sc app-uMaf}), and let $M$ be an edge-removal meta-step on $I$ producing instance $I'$. If $M$ keeps ratio $r$, then $\mbox{Opt}(I') - \mbox{Opt}(I) \leq \frac{r-1}{r}|\widetilde{E_M^1}|$.

\begin{proof}
Let $F$ be a fixed MAF for $I=(F_1,F_2,\ldots,F_m)$. According to Definition-R, there exists an edge-set $E_M^{1}$ that no edge in $E_M \setminus E_M^{1}$ is in any agreement forest for $(F_1 \setminus E_M^1,F_2,\ldots,F_m)$. Thus, instances $I'$ and $(F_1 \setminus E_M^1,F_2,\ldots,F_m)$ have the same collection of solutions. Because $F_1 \setminus E_M^1 = F_1 \setminus \widetilde{E_M^1}$, so $I'$ and $(F_1 \setminus \widetilde{E_M^1},F_2,\ldots,F_m)$ also have the same collection of solutions. $\mbox{Opt}(I')$ is the same as the optimal order for $(F_1 \setminus \widetilde{E_M^1},F_2,\ldots,F_m)$.

Since $F$ is an agreement forest for $(F_1,F_2,\ldots,F_m)$, in order to construct an agreement forest for $(F_1 \setminus \widetilde{E_M^1},F_2,\ldots,F_m)$ by removing edges from $F$, we just need removing the edges from $F$ that are not in $F_1 \setminus \widetilde{E_M^1}$ to make the new $F$ be a subforest of $F_1 \setminus \widetilde{E_M^1}$. Here, we denote by $E_1$ the subset of $\widetilde{E_M^1}$ in which the edges are in $F$, and denote by $E_2$ the subset of $\widetilde{E_M^1}$ in which the edges are not in $F$. Obviously, $F \setminus E_1$ is an agreement forest for $(F_1 \setminus \widetilde{E_M^1},F_2,\ldots,F_m)$. In the following, we analyze the order of $F \setminus E_1$ detailly.

According to Definition-R, we have that for $F$, there exists an ee-set $E_M^{1,F}$ of $F_1$, $E_M^{1,F} \subseteq E_M^1$, $|E_M^{1,F}| \geq |\widetilde{E_M^1}| / r$, and no edge in $E_M^{1,F}$ is in $F$. Note that we can easily get an essential subset of $E_M^{1}$ that contains $E_M^{1,F}$, thus, we can assume that $E_M^{1,F} \subseteq \widetilde{E_M^1}$. Therefore, we have that $E_M^{1,F} \subseteq E_2$ and $|E_2| \geq |\widetilde{E_M^1}|/r$. Because $\widetilde{E_M^1} \setminus E_2 = E_1$, so $|E_1| = |\widetilde{E_M^1} \setminus E_2| \leq \frac{r-1}{r}|\widetilde{E_M^1}|$. Therefore, $\mbox{Ord}(F \setminus E_1) \leq \mbox{Ord}(F) + \frac{r-1}{r}|\widetilde{E_M^1}|$.

Since $\mbox{Opt}(I') \leq \mbox{Ord}(F \setminus E_1)$ and $\mbox{Opt}(I) = \mbox{Ord}(F)$, we have $\mbox{Opt}(I') - \mbox{Opt}(I) \leq \frac{r-1}{r}|\widetilde{E_M^1}|$.
\end{proof}
\end{lemma}

\begin{theorem}
\label{ratio}
Let $I = (F_1,F_2,\ldots,F_m)$ be an instance of {\sc app-Maf} ({\sc app-rMaf} or {\sc app-uMaf}), and let $t \geq 1$ be an arbitrary real number. If each meta-step in Step 3 of algorithm Alg-{\sc app-Maf} keeps ratio not greater than $t$ and that the algorithm Alg-{\sc app-Maf} halts on the instance $I$, then the output of Alg-{\sc app-Maf} is an agreement forest for $I$ whose order is at most $t$ times the optimal value for $I$.

\begin{proof}
Suppose the sequence of meta-steps in the algorithm is $S = \{M_1,M_2,\ldots,M_h\}$, where for each $i$, $1\leq i \leq h$, meta-step $M_i$ removes an edge-set $E_{M_i}$ from the instance
$I_i=(F_{1,i},F_{2,i},\ldots,F_{m,i})$ produces an instance $I_{i+1}=(F_{1,i+1},F_{2,i+1},\ldots,F_{m,i+1})$. By the judgement condition in Step 3, we can make sure that $F_{1,h+1}$ is a subforest of $F_{2,h+1}, \ldots, F_{m,h+1}$. Thus, $F_{1,h+1}$ is an MAF for $I_{h+1}$, Ord$(F_{1,h+1})=\mbox{Opt}(I_{h+1})$.

Because each meta-step $M_i$ in $S$, $1\leq i \leq h$, removes certain edges from forests in $I_i$, so every agreement forest for $I_{i+1}$ is also an agreement forest for $I_i$. Therefore, the forest $F_{1,h+1}$ returned by Step 4 is also an agreement forest for the original input instance $I_1$.

For each meta-step $M_i \in S$, $1 \leq i \leq h$, by Lemma~\ref{equation}, there is $(\mbox{Opt}(I_{i+1})-\mbox{Opt}(I_{i})) \leq \frac{t-1}{t}|\widetilde{E_M^1}|$. Note that, $|\widetilde{E_M^1}| = \mbox{Ord}(F_{1,i+1}) - \mbox{Ord}(F_{1,i})$. Therefore, for each meta-step $M_i \in S$, we have the inequality $(\mbox{Opt}(I_{i+1})-\mbox{Opt}(I_{i})) \leq \frac{t-1}{t}(\mbox{Ord}(F_{1,i+1})-\mbox{Ord}(F_{1,i}))$.

Then, we add up these inequalities for all meta-steps in $S$, and get $(\mbox{Opt}(I_{h+1}) - \mbox{Opt}(I_{1})) \leq \frac{t-1}{t}(\mbox{Ord}(F_{1,h+1}) - \mbox{Ord}(F_{1,1}))$, where $\mbox{Opt}(I_{h+1}) = \mbox{Ord}(F_{1,h+1})$. From this, we can easily get $\mbox{Opt}(I_{h+1}) \leq t*\mbox{Opt}(I_{1})$, which proves the theorem.
\end{proof}
\end{theorem}

\subsection{Approximation Algorithm for {\sc app-rMaf}}

We develop meta-steps for {\sc app-rMaf} in this subsection. Thus, all X-forests considered in this subsection are rooted. As given in the algorithm Alg-{\sc app-Maf} (see Figure~\ref{alg-app}), for each execution of Step 3 in the algorithm, we are given a fixed integer $i > 1$ and an instance $I = (F_1,F_2,\ldots,F_m)$ of {\sc app-rMaf}, which is a collection of rooted X-forests, with $F_1$ is a subforest of  $F_2,\ldots,F_{i-1}$, and, as long as $F_1 \neq F_i$, meta-steps are applied on $F_1$ and $F_i$. In the following, we show how these meta-steps are constructed based on different structures of $F_1$ and $F_i$.

Let $F^*$ be a fixed MAF for $I = (F_1,F_2,\ldots,F_m)$, and let $F'$ be a maximal-AF for $F_1$ and $F_i$ that contains $F^*$. Since $F^*$ is a subforest of $F'$, if an edge $e$ of $F_1$ is not in $F'$, then $e$ is also not in $F^*$.

An execution of Reduction Rule 1 on $F_1$ and $F_i$ can be regarded as an edge-removal meta-step. By Lemma~\ref{Reduction Rule 1} and Remark 2 of Definition-R, we can easily get the following lemma.

\begin{lemma}
\label{Reduction Rule 1 safe}
Reduction Rule 1 is safe.
\end{lemma}

By Lemma~\ref{simpleroot}, if $F_i$ has no MSS, then $F_1$ and $F_i$ have an unique MAF $F$, which either is isomorphic to $F_i$ or consists of a collection of single-vertex trees. And by a series of executions of Reduction Rule 1 on $F_1$ and $F_i$, there is $F_1 = F_i = F$, which satisfies the judgement condition of Step 3 in algorithm Alg-{\sc app-Maf}. Therefore, in the following discussion, we will assume that $F_i$ has an MSS $S$. Note that the instances in our discussion are strongly reducible, so none of labels in $S$ is a single-vertex tree in $F_1$. W.l.o.g., we will assume that labels $a$ and $b$ belong to $S$.

\smallskip

\noindent{\bf Case 1.} All labels in $S$ consist an MSS in $F_1$.

\smallskip

\noindent{\bf Meta-step 1.} Group all labels in $S$ and their parent into an un-decomposable structure, and mark the unit with the same label in $F_1$ and $F_i$.

\smallskip

The implementation of Meta-step 1 is the same as that of Step 1 for {\sc para-rMaf}.

This meta-step can be regarded as a special meta-step that does not remove any edges in the instance. Instead, it groups certain structures in some X-forests into un-decomposable units. Using the notation in Definition-R, $E_M = \emptyset$. And, we have

\begin{lemma}
\label{Meta-step 1}
Meta-step 1 is safe.
\end{lemma}

This meta-step may lead a subtle problem in the following discussion. Because Meta-step 1 changes the label-sets of $F_1$ and $F_i$, so the label-sets of $F_1$ and $F_i$ are different from the label-sets of the other forests in the instance. Thus, there is ambiguity of the sentence ``there exists an MAF for the new instance".

Note that this operation is just simply for notational convenience. We still can construct an MAF for the new instance if we ``expand" these combined leaves in $F_1$ and $F_i$. Therefore, in the following discussion, we can simply say that there exists an MAF $F^*$ for the new instance, although the label-sets of the forests in the instance are different.

Note again that in the MAF $F^*$ for the instance, there maybe no such an un-decomposable structure, because some labels in $S$ maybe in different connected components in $F^*$. However, the maximal-AF $F'$ for $F_1$ and $F_i$ that contains $F^*$ must have such an un-decomposable structure. Therefore, in the following discussion, we can assume that the maximal-AF $F'$ for $F_1$ and $F_i$ that contains $F^*$ has been applied all possible ``group" operations so that $F_1$, $F_i$, and $F'$ have the same new label-set. Similarly, even though $F^*$ and $F'$ have different label-sets, we can also simply say that $F'$ contains $F^*$, because we just need to expand these combined leaves in $F'$.

Let $\widehat{F'}$ be the forest that getting by expanding these combined leaves in $F'$. Obviously, $\widehat{F'}$ and $F^*$ have the same label-set, and $F^*$ is a subforest of $\widehat{F'}$. Therefore, for any edge $e$ that in $F_1$ or $F_i$, if $e$ is not in $F'$, then $e$ is not in $\widehat{F'}$. Thus, $e$ is also not in $F^*$.

\smallskip

\noindent{\bf Case 2.} All labels in $S$ are siblings in $F_1$.

\smallskip

Let $p_1$ be the common parent of $S$ in $F_1$. In this case, there must exist a vertex $v$ in $F_1$ which is a child of $p_1$ but not belongs to $S$ ($v$ maybe a non-leaf). Let $e$ be the edge that between $p_1$ and $v$ in $F_1$.

\smallskip

\noindent{\bf Meta-step 2.} Remove the edges that incident to $a$ and $b$ in $F_1$ and $F_i$, and remove edge $e$.

\begin{lemma}
\label{Meta-step 2}
Meta-step 2 keeps ratio $3$.

\begin{proof}
Let $e_a$ and $e_b$ be the edges that incident to $a$ and $b$ in $F_1$, respectively, and let $e'_a$ and $e'_b$ be the edges that incident to $a$ and $b$ in $F_i$, respectively. Using the notations in Definition-R, we have $E_M = \{e_a,e_b,e'_a,e'_b,e\}$ and $E_M^1 = \{e_a,e_b,e\}$.

In the $X$-forest $F_1 \setminus E_M^1$, labels $a$ and $b$ are single-vertex trees. Thus, every maximal-AF for $F_1 \setminus E_M^1$ and $F_i$ have $a$ and $b$ as single-vertex trees, no edges in $\{e'_a,e'_b\}$ can be in any maximal-AF for $F_1 \setminus E_M^1$ and $F_i$. Because every agreement forest for $(F_1 \setminus E_M^1,F_2,\ldots,F_m)$ must be a subforest of a maximal-AF for $F_1 \setminus E_M^1$ and $F_i$, so no edges in $\{e'_a,e'_b\}$ can be in any agreement forest for $(F_1 \setminus E_M^1,F_2,\ldots,F_m)$.

There are three situations for $a$ and $b$ in the maximal-AF $F'$ for $F_1$ and $F_i$ which contains $F^*$.

Situation 1. $a$ is a single-vertex tree in $F'$. Thus, $e_a$ is not in $F'$, and $e_a$ is also not in $F^*$. Therefore, we can pick $\{e_a\}$ as the set $E_M^{1,F^*}$, which satisfies: $E_M^{1,F^*} \subseteq E_M^{1}$, $|E_M^{1,F^*}| \geq |\widetilde{E_M^{1}}|/3$. Note that $|\widetilde{E_M^{1}}|$ is not greater than $|E_M^1| = 3$. Moreover, since $F_1$ is irreducible and $a$ is not a single-vertex tree in $F_1$, the set $E_M^{1,F^*}$ is an ee-set of $F_1$. Therefore, for the agreement forest $F^*$, the set $E_M^{1,F^*}$ satisfies all conditions in Definition-R to make meta-step 2 to keep a ratio $3$.

Situation 2. $b$ is a single-vertex tree in $F'$. Thus, $e_b$ is not in $F'$, and $e_b$ is also not in $F^*$. Then similarly we let $E_M^{1,F^*} = \{e_b\}$ and can verify that for the agreement forest $F^*$, the set $E_M^{1,F^*}$ satisfies all conditions in Definition-R to make meta-step 2 to keep a ratio $3$.

Situation 3. $a$ and $b$ are siblings in $F'$. Then, by Lemma~\ref{allsiblings-rooted}, the labels of $S$ consist an MSS in $F'$. In order to make the labels of $S$ consist an MSS in $F_1$, edge $e$ should be removed. That is, in this situation, edge $e$ is not in $F'$, and $e$ is also not in $F^*$. Thus, in this situation, we let $E_M^{1,F^*} = \{e\}$, and verify easily that for the agreement forest $F^*$, the set $E_M^{1,F^*}$ satisfies all conditions in Definition-R to make meta-step 2 to keep a ratio $3$.

This verifies that the set $E_M^1$ satisfies all conditions in Definition-R to make meta-step 2 to keep a ratio $3$. Thus, Meta-step 2 keeps ratio $3$.
\end{proof}
\end{lemma}

\noindent{\bf Case 3.} Some labels in $S$ are not siblings in $F_1$.

\smallskip

W.l.o.g., we assume $a$ and $b$ are not siblings in $F_1$.

\smallskip

\noindent{\bf Subcase 3.1.} $a$ and $b$ are in different connected components in $F_1$.

\smallskip

\noindent{\bf Meta-step 3.1.} Remove the edges incident to $a$ and $b$ in both $F_1$ and $F_i$.

\begin{lemma}
\label{Meta-step 3.1}
Meta-step 3.1 keeps ratio $2$.

\begin{proof}
Let $e_a$ and $e_b$ be the edges that incident to $a$ and $b$ in $F_1$, respectively, and let $e'_a$ and $e'_b$ be the edges that incident to $a$ and $b$ in $F_i$, respectively. Using the notations in Definition-R, we have $E_M = \{e_a,e_b,e'_a,e'_b\}$ and $E_M^1 = \{e_a,e_b\}$.

In the $X$-forest $F_1 \setminus E_M^1$, labels $a$ and $b$ are single-vertex trees. Thus, every maximal-AF for $F_1 \setminus E_M^1$ and $F_i$ have $a$ and $b$ as single-vertex trees, no edges in $\{e'_a,e'_b\}$ can be in any maximal-AF for $F_1 \setminus E_M^1$ and $F_i$. Because every agreement forest for $(F_1 \setminus E_M^1,F_2,\ldots,F_m)$ must be a subforest of a maximal-AF for $F_1 \setminus E_M^1$ and $F_i$, so no edges in $\{e'_a,e'_b\}$ can be in any agreement forest for $(F_1 \setminus E_M^1,F_2,\ldots,F_m)$.

There are two situations for $a$ and $b$ in the maximal-AF $F'$ for $F_1$ and $F_i$ which contains $F^*$.

Situation 1. $a$ is a single-vertex tree in $F'$. Thus, $e_a$ is not in $F'$, and $e_a$ is also not in $F^*$. Therefore, we can pick $\{e_a\}$ as the set $E_M^{1,F^*}$, which satisfies: $E_M^{1,F^*} \subseteq E_M^{1}$, $|E_M^{1,F^*}| \geq |\widetilde{E_M^{1}}|/2$. Moreover, since $F_1$ is irreducible and $a$ is not a single-vertex tree in $F_1$, the set $E_M^{1,F^*}$ is an ee-set of $F_1$. Therefore, for the agreement forest $F^*$, the set $E_M^{1,F^*}$ satisfies all conditions in Definition-R to make meta-step 3.1 to keep a ratio $2$.

Situation 2. $b$ is a single-vertex tree in $F'$. Thus, $e_b$ is not in $F'$, and $e_b$ is also not in $F^*$. Then similarly we let $E_M^{1,F^*} = \{e_b\}$ and can verify that for the agreement forest $F^*$, the set $E_M^{1,F^*}$ satisfies all conditions in Definition-R to make meta-step 3.1 to keep a ratio $2$.

This verifies that the set $E_M^1$ satisfies all conditions in Definition-R to make meta-step 3.1 to keep a ratio $2$. Thus, Meta-step 3.1 keeps ratio $2$.
\end{proof}
\end{lemma}

\noindent{\bf Subcase 3.2.} $a$ and $b$ are in the same connected component in $F_1$.

\smallskip

Let $P = \{a,c_1,\ldots,c_r,b\}$ be the path that connects $a$ and $b$ in $F_1$, and let $c_h$ be the least common ancestor of $a$ and $b$ in $F_1$, $1 \leq h \leq r$. Let $E_p$ be the edge set that contains all edge that incident to $c_i$, $1 \leq i \leq r$, $i \neq h$, but not on the path $P$, and let $e$ be an arbitrary edge of $E_p$.

\smallskip

\noindent{\bf Meta-step 3.2.} Remove the edges incident to $a$ and $b$ in both $F_1$ and $F_i$, and remove edge $e$.

\begin{lemma}
\label{Meta-step 3.2}
Meta-step 3.2 keeps ratio $3$.

\begin{proof}
Let $e_a$ and $e_b$ be the edges that incident to $a$ and $b$ in $F_1$, respectively, and let $e'_a$ and $e'_b$ be the edges that incident to $a$ and $b$ in $F_i$, respectively. Using the notations in Definition-R, we have $E_M = \{e_a,e_b,e'_a,e'_b,e\}$ and $E_M^1 = \{e_a,e_b,e\}$.

In the $X$-forest $F_1 \setminus E_M^1$, labels $a$ and $b$ are single-vertex trees. Thus, every maximal-AF for $F_1 \setminus E_M^1$ and $F_i$ have $a$ and $b$ as single-vertex trees, no edges in $\{e'_a,e'_b\}$ can be in any maximal-AF for $F_1 \setminus E_M^1$ and $F_i$. Because every agreement forest for $(F_1 \setminus E_M^1,F_2,\ldots,F_m)$ must be a subforest of a maximal-AF for $F_1 \setminus E_M^1$ and $F_i$, so no edges in $\{e'_a,e'_b\}$ can be in any agreement forest for $(F_1 \setminus E_M^1,F_2,\ldots,F_m)$.

There are three situations for $a$ and $b$ in the maximal-AF $F'$ for $F_1$ and $F_i$ which contains $F^*$.

Situation 1. $a$ is a single-vertex tree in $F'$. Thus, $e_a$ is not in $F'$, and $e_a$ is also not in $F^*$. Therefore, we can pick $\{e_a\}$ as the set $E_M^{1,F^*}$, which satisfies: $E_M^{1,F^*} \subseteq E_M^{1}$, $|E_M^{1,F^*}| \geq |\widetilde{E_M^{1}}|/3$. Moreover, since $F_1$ is irreducible and $a$ is not a single-vertex tree in $F_1$, the set $E_M^{1,F^*}$ is an ee-set of $F_1$. Therefore, for the agreement forest $F^*$, the set $E_M^{1,F^*}$ satisfies all conditions in Definition-R to make meta-step 3.2 to keep a ratio $3$.

Situation 2. $b$ is a single-vertex tree in $F'$. Thus, $e_b$ is not in $F'$, and $e_b$ is also not in $F^*$. Then similarly we let $E_M^{1,F^*} = \{e_b\}$ and can verify that for the agreement forest $F^*$, the set $E_M^{1,F^*}$ satisfies all conditions in Definition-R to make meta-step 3.2 to keep a ratio $3$.

Situation 3. $a$ and $b$ are siblings in $F'$. In order to make labels $a$ and $b$ be siblings in $F_1$, all the edges in $E_p$ should be removed. That is, in this situation, all the edges in $E_p$ are not in $F'$, so all the edges in $E_p$ are not in $F^*$. Here, we just remove the edge $e$ in $E_p$, so we let $E_M^{1,F^*} = \{e\}$, and verify easily that for the agreement forest $F^*$, the set $E_M^{1,F^*}$ satisfies all conditions in Definition-R to make meta-step 3.2 to keep a ratio $3$.

This verifies that the set $E_M^1$ satisfies all conditions in Definition-R to make meta-step 3.2 to keep a ratio $3$. Thus, Meta-step 3.2 keeps ratio $3$.
\end{proof}
\end{lemma}

\smallskip

For two rooted X-forests $F_1$ and $F_i$, if we iteratively apply the above process based on the cases, then the process will end up that $F_1$ is isomorphic to $F_i$. (To here, we also ``expand" these combined labels in $F_1$ and $F_i$.)

Now, combining the general framework given in Figure~\ref{alg-app} and the meta-steps given above, we are ready to present our approximation algorithm for {\sc app-rMaf}.

\begin{figure}[gpath3]
\setbox4=\vbox{\hsize28pc \noindent\strut \begin{quote}
\vspace*{-6mm} \footnotesize

{\bf Algorithm} Alg-{\sc app-rMaf}$(F_1, F_2, \ldots, F_m)$\\
Input: a collection $\{F_1, F_2, \ldots, F_m\}$ of rooted general $X$-forests, $m \geq 1$\\
Output: an agreement forest $F^*$ for $\{F_1, F_2, \ldots, F_m\}$\\
1. {\bf if} $m=1$ {\bf then} return $F_1$;\\
2. {\bf for} $i = 2$ to $m$ {\bf do}\\
3. \hspace*{3mm} {\bf while} $F_1 \neq F_i$\\
  \hspace*{18mm} {\bf apply} Reduction Rule 1 on $F_1$ and $F_i$ if possible;\\
  \hspace*{18mm} {\bf if} $F_i$ contains an MSS, then let $S$ be an MSS of $F_i$;\\
  \hspace*{18mm} {\bf switch}\\
 \hspace*{25mm} Case 1:\hspace*{3mm} apply Meta-step 1;\\
 \hspace*{25mm} Case 2:\hspace*{3mm} apply Meta-step 2;\\
 \hspace*{25mm} Case 3.1: apply Meta-step 3.1;\\
 \hspace*{25mm} Case 3.2: apply Meta-step 3.2;\\
4. {\bf return} $F_1$.
\end{quote}
\vspace*{-6mm} \strut} $$\boxit{\box4}$$ \vspace*{-9mm}
\caption{Algorithm for the {\sc app-rMaf} problem}
\label{algrooted-app}
\end{figure}

\begin{theorem}
\label{alg-app-rMaf-ratio}
Algorithm Alg-{\sc app-rMaf} is $3$-approximation algorithm for the {\sc app-rMaf} problem that runs in time $O(mn^2)$, where $n$ is the size of the label-set $X$ and $m$ is the number of forests in the input instance.

\begin{proof}
By Lemmas~\ref{Meta-step 1},~\ref{Meta-step 2},~\ref{Meta-step 3.1}, and~\ref{Meta-step 3.2}, each of the meta-steps keeps a ratio bounded by $3$. By Theorem~\ref{ratio}, if the algorithm Alg-{\sc app-Maf} uses these meta-steps in Step 3, and halts on an instance $I$ of {\sc app-rMaf}, then the algorithm Alg-{\sc app-rMaf} produces an agreement forest for the instance $I$ whose order is bounded by $3$ times the optimal value for $I$. Therefore, to show that the algorithm Alg-{\sc app-rMaf} is a $3$-approximation algorithm for the {\sc app-rMaf} problem, it suffices to show that on any instance $I$ of {\sc app-rMaf}, the algorithm Alg-{\sc app-rMaf} runs in time $O(nm \log n)$.

By the above discussion, as long as $F_1 \neq F_i$, at least one of the above meta-steps is applicable. Let $n=|X|$. Then, the number of vertices plus the number of edges in an X-forest is $O(n)$. By the algorithm Alg-{\sc app-rMaf}, a meta-step in Step 3 is applied on $F_1$ and $F_i$ only when $F_1 \neq F_i$. Under the condition $F_1 \neq F_i$, it is easy to verify that each of the meta-steps 2, 3.1, 3.2, and Reduction Rule 1 removes at least one edge in $F_1 \cup F_i$. Therefore, the total number of times these meta-steps can be applied is bounded by $O(n)$.

Now consider meta-step 1. Initially, each vertex in $F_1$
and $F_i$ is an un-decomposable unit. Thus, the total number of un-decomposable units in $F_1 \cup F_i$ is $O(n)$. Each application of meta-step 1 groups three un-decomposable units into a single un-decomposable unit, in each of $F_1$ and $F_i$. Therefore, meta-step 1 can be applied at most $O(n)$ times.

Summarizing the above discussion, we conclude that if the algorithm Alg-{\sc app-rMaf} uses these meta-steps in Step 3, then the number of times meta-steps are applied in each execution of Step 3 is $O(n)$. Moreover, it is not very difficult to see that with careful implementation of the data structure representing X-forests, the running time of each of the meta-steps, can be bounded by $O(n)$. Therefore, the running time of the algorithm is $O(mn^2)$, where $n=|X|$ and $m$ is the number of X-forests in the input instance.
\end{proof}
\end{theorem}

\subsection{Approximation Algorithm for {\sc app-uMaf}}

In this subsection, we develop meta-steps for {\sc app-uMaf}. Let $F^*$ be a fixed MAF for the instance $I=(F_1,F_2,\ldots,F_m)$ of {\sc app-uMaf}, and let $F'$ be a maximal-AF for $F_1$ and $F_i$ that contains $F^*$.

By Lemma~\ref{simpleunroot}, if $F_i$ has no MSS, then $F_1$ and $F_i$ have an unique MAF, which consists of a collection of single-vertex trees. And by a series of executions of Reduction Rule 1 on $F_1$ and $F_i$, we can get that $F_1 = F_i$, which satisfies the judgement condition of Step 3 in algorithm Alg-{\sc app-Maf}. Therefore, in the following discussion, we will assume that $F_i$ has an MSS $S$. Note that none of labels in $S$ is a single-vertex tree in $F_1$. W.l.o.g., we will also assume that labels $a$ and $b$ belong to $S$.

\smallskip

\noindent{\bf Case 1.} All labels in $S$ consist an MSS in $F_1$.

\smallskip

\noindent{\bf Meta-step 1.} Group all labels in $S$ (and their common neighbor if $S$ is not the label set of a single-edge tree) into an un-decomposable structure, and mark the unit with the same label in $F_1$ and $F_2$.

\smallskip

The implementation of Meta-step 1 is the same as that of Step 1 for {\sc para-uMaf}. This meta-step also can be regarded as a special meta-step that does not remove any edges in the instance. And, we have

\begin{lemma}
\label{unrooted Meta-step 1}
Meta-step 1 is safe.
\end{lemma}

\noindent{\bf Case 2.} All labels in $S$ are siblings in $F_1$.

\smallskip

Obviously, the common neighbor $p$ of $S$ in $F_1$ has degree greater than $|X|+1$. Let $e_1$ and $e_2$ be two arbitrarily edges in $F_1$ that incident to $p$ but not incident to the labels in $S$.

\smallskip

\noindent{\bf Meta-step 2.} Remove the edges that incident to $a$ and $b$ in both $F_1$ and $F_2$, and remove edges $e_1$ and $e_2$.

\begin{lemma}
\label{unrooted Meta-step 2}
Meta-step 2 keeps ratio $4$.

\begin{proof}
Let $e_a$ and $e_b$ be the edges that incident to $a$ and $b$ in $F_1$, respectively, and let $e'_a$ and $e'_b$ be the edges that incident to $a$ and $b$ in $F_i$, respectively. Using the notations in Definition-R, we have $E_M = \{e_a,e_b,e'_a,e'_b,e_1,e_2\}$ and $E_M^1 = \{e_a,e_b,e_1,e_2\}$.

In the $X$-forest $F_1 \setminus E_M^1$, labels $a$ and $b$ are single-vertex trees. Thus, every maximal-AF for $F_1 \setminus E_M^1$ and $F_i$ have $a$ and $b$ as single-vertex trees, no edges in $\{e'_a,e'_b\}$ can be in any maximal-AF for $F_1 \setminus E_M^1$ and $F_i$. Because every agreement forest for $(F_1 \setminus E_M^1,F_2,\ldots,F_m)$ must be a subforest of a maximal-AF for $F_1 \setminus E_M^1$ and $F_i$, so no edges in $\{e'_a,e'_b\}$ can be in any agreement forest for $(F_1 \setminus E_M^1,F_2,\ldots,F_m)$.

There are three situations for $a$ and $b$ in the maximal-AF $F'$ for $F_1$ and $F_i$ which contains $F^*$.

Situation 1. $a$ is a single-vertex tree in $F'$. Thus, $e_a$ is not in $F'$, and $e_a$ is also not in $F^*$. Therefore, we can pick $\{e_a\}$ as the set $E_M^{1,F^*}$, which satisfies: $E_M^{1,F^*} \subseteq E_M^{1}$, $|E_M^{1,F^*}| \geq |\widetilde{E_M^{1}}|/4$. Note that $|\widetilde{E_M^{1}}|$ is not greater than $|E_M^1| = 4$. Moreover, since $F_1$ is irreducible and $a$ is not a single-vertex tree in $F_1$, the set $E_M^{1,F^*}$ is an ee-set of $F_1$. Therefore, for the agreement forest $F^*$, the set $E_M^{1,F^*}$ satisfies all conditions in Definition-R to make meta-step 2 to keep a ratio $4$.

Situation 2. $b$ is a single-vertex tree in $F'$. Thus, $e_b$ is not in $F'$, and $e_b$ is also not in $F^*$. Then similarly we let $E_M^{1,F^*} = \{e_b\}$ and can verify that for the agreement forest $F^*$, the set $E_M^{1,F^*}$ satisfies all conditions in Definition-R to make meta-step 2 to keep a ratio $4$.

Situation 3. $a$ and $b$ are siblings in $F'$. By Lemma~\ref{allsiblings-unrooted}, all labels of $S$ are siblings in $F'$. Because the labels of $S$ consist an MSS in $F_i$, so the labels of $S$ consist an MSS in $F'$. In order to make the labels of $S$ consist an MSS in $F_1$, the degree of $p$ cannot be greater than $|X|+1$. That is, at least one of edges $e_1$ and $e_2$ in $F_1$ should be removed. If $e_1$ is not in $F'$, then $e_1$ is also not in $F^*$. Let $E_M^{1,F^*} = \{e_1\}$, we can easily verify that for the agreement forest $F^*$, the set $E_M^{1,F^*}$ satisfies all conditions in Definition-R to make meta-step 2 to keep a ratio $4$. The same argument is applied to the case that $e_2$ is not in $F'$.

This verifies that the set $E_M^1$ satisfies all conditions in Definition-R to make meta-step 2 to keep a ratio $4$. Thus, Meta-step 2 keeps ratio $4$.
\end{proof}
\end{lemma}

\noindent{\bf Case 3.} Some labels in $S$ are not siblings in $F_1$.

\smallskip

W.l.o.g., we will assume that labels $a$ and $b$ are not siblings in $F_1$.

\smallskip

\noindent{\bf Subcase 3.1.} $a$ and $b$ are in different connected components in $F_1$.

\smallskip

\noindent{\bf Meta-step 3.1.} Remove the edges incident to $a$ and $b$ in both $F_1$ and $F_i$.

\begin{lemma}
\label{unrooted Meta-step 3.1.}
Meta-step 3.1 keeps ratio $2$.

\begin{proof}
Let $e_a$ and $e_b$ be the edges that incident to $a$ and $b$ in $F_1$, respectively, and let $e'_a$ and $e'_b$ be the edges that incident to $a$ and $b$ in $F_i$, respectively. Using the notations in Definition-R, we have $E_M = \{e_a,e_b,e'_a,e'_b\}$ and $E_M^1 = \{e_a,e_b\}$.

In the $X$-forest $F_1 \setminus E_M^1$, labels $a$ and $b$ are single-vertex trees. Thus, every maximal-AF for $F_1 \setminus E_M^1$ and $F_i$ have $a$ and $b$ as single-vertex trees, no edges in $\{e'_a,e'_b\}$ can be in any maximal-AF for $F_1 \setminus E_M^1$ and $F_i$. Because every agreement forest for $(F_1 \setminus E_M^1,F_2,\ldots,F_m)$ must be a subforest of a maximal-AF for $F_1 \setminus E_M^1$ and $F_i$, so no edges in $\{e'_a,e'_b\}$ can be in any agreement forest for $(F_1 \setminus E_M^1,F_2,\ldots,F_m)$.

There are two situations for $a$ and $b$ in the maximal-AF $F'$ for $F_1$ and $F_i$ which contains $F^*$.

Situation 1. $a$ is a single-vertex tree in $F'$. Thus, $e_a$ is not in $F'$, so $e_a$ also is not in $F^*$. Therefore, we can pick $\{e_a\}$ as the set $E_M^{1,F^*}$, which satisfies: $E_M^{1,F^*} \subseteq E_M^{1}$, $|E_M^{1,F^*}| \geq |\widetilde{E_M^{1}}|/2$. Moreover, since $F_1$ is irreducible and $a$ is not a single-vertex tree in $F_1$, the set $E_M^{1,F^*}$ is an ee-set of $F_1$. Therefore, for the agreement forest $F^*$, the set $E_M^{1,F^*}$ satisfies all conditions in Definition-R to make meta-step 3.1 to keep a ratio $2$.

Situation 2. $b$ is a single-vertex tree in $F'$. Thus, $e_b$ is not in $F'$, so $e_b$ also is not in $F^*$. Then similarly we let $E_M^{1,F^*} = \{e_b\}$ and can verify that for the agreement forest $F^*$, the set $E_M^{1,F^*}$ satisfies all conditions in Definition-R to make meta-step 3.1 to keep a ratio $2$.

This verifies that the set $E_M^1$ satisfies all conditions in Definition-R to make meta-step 3.1 to keep a ratio $2$. Thus, Meta-step 3.1 keeps ratio $2$.
\end{proof}
\end{lemma}

\noindent{\bf Subcase 3.2.} $a$ and $b$ are in the same connected component in $F_1$.

\smallskip

Let $P = \{a,c_1,\ldots,c_r,b\}$ be the path that connects $a$ and $b$ in $F_1$. And let $E_p$ be the edge set that contains all edge that incident to $c_i$, $1 \leq i \leq r$, but not on the path $P$. Obviously, $|E_p| \geq 2$. Let $e_1$ and $e_2$ be two arbitrary edges in $E_p$.

\smallskip

\noindent{\bf Meta-step 3.2.} Remove the edges that incident to $a$ and $b$ in both $F_1$ and $F_i$, and remove edges $e_1$ and $e_2$ in $F_1$.

\begin{lemma}
\label{Meta-step 3.2.}
Meta-step 3.2 keeps ratio $4$.

\begin{proof}
Let $e_a$ and $e_b$ be the edges that incident to $a$ and $b$ in $F_1$, respectively, and let $e'_a$ and $e'_b$ be the edges that incident to $a$ and $b$ in $F_i$, respectively. Using the notations in Definition-R, we have $E_M = \{e_a,e_b,e'_a,e'_b,e_1,e_2\}$ and $E_M^1 = \{e_a,e_b,e_1,e_2\}$.

In the $X$-forest $F_1 \setminus E_M^1$, labels $a$ and $b$ are single-vertex trees. Thus, every maximal-AF for $F_1 \setminus E_M^1$ and $F_i$ have $a$ and $b$ as single-vertex trees, no edges in $\{e'_a,e'_b\}$ can be in any maximal-AF for $F_1 \setminus E_M^1$ and $F_i$. Because every agreement forest for $(F_1 \setminus E_M^1,F_2,\ldots,F_m)$ must be a subforest of a maximal-AF for $F_1 \setminus E_M^1$ and $F_i$, so no edges in $\{e'_a,e'_b\}$ can be in any agreement forest for $(F_1 \setminus E_M^1,F_2,\ldots,F_m)$.

There are three situations for $a$ and $b$ in the maximal-AF $F'$ for $F_1$ and $F_i$ which contains $F^*$.

Situation 1. $a$ is a single-vertex tree in $F'$. Thus, $e_a$ is not in $F'$, and $e_a$ is also not in $F^*$. Therefore, we can pick $\{e_a\}$ as the set $E_M^{1,F^*}$, which satisfies: $E_M^{1,F^*} \subseteq E_M^{1}$, $|E_M^{1,F^*}| \geq |\widetilde{E_M^{1}}|/4$. Note that $|\widetilde{E_M^{1}}|$ is always less than $|E_M^1| = 4$. Moreover, since $F_1$ is irreducible and $a$ is not a single-vertex tree in $F_1$, the set $E_M^{1,F^*}$ is an ee-set of $F_1$. Therefore, for the agreement forest $F^*$, the set $E_M^{1,F^*}$ satisfies all conditions in Definition-R to make meta-step 3.2 to keep a ratio $4$.

Situation 2. $b$ is a single-vertex tree in $F'$. Thus, $e_b$ is not in $F'$, and $e_b$ is also not in $F^*$. Then similarly we let $E_M^{1,F^*} = \{e_b\}$ and can verify that for the agreement forest $F^*$, the set $E_M^{1,F^*}$ satisfies all conditions in Definition-R to make meta-step 3.2 to keep a ratio $4$.

Situation 3. $a$ and $b$ are siblings in $F'$. Thus, $a$ and $b$ either are connected by an edge or have a common neighbor in $F'$.
If $a$ and $b$ are connected by an edge in $F'$, then all edges in $E_p$ should be removed. Here, we just remove edges $e_1$ and $e_2$, and let $E_M^{1,F^*} = \{e_1,e_2\}$, we can easily verify that for the agreement forest $F^*$, the set $E_M^{1,F^*}$ satisfies all conditions in Definition-R to make meta-step 3.2 to keep a ratio $4$. If $a$ and $b$ have a common neighbor in $F'$, then at least one of $e_1$ and $e_2$ should be removed. If $e_1$ is not in $F'$, then $e_1$ is also not in $F^*$. Let $E_M^{1,F^*} = \{e_1\}$, we can verify that for the agreement forest $F^*$, the set $E_M^{1,F^*}$ satisfies all conditions in Definition-R to make meta-step 3.2 to keep a ratio $4$. The same argument is applied to the case that $e_2$ is not in $F'$.

This verifies that the set $E_M^1$ satisfies all conditions in Definition-R to make meta-step 3.2 to keep a ratio $4$. Thus, Meta-step 3.2 keeps ratio $4$.
\end{proof}
\end{lemma}

For two unrooted general X-forests $F_1$ and $F_i$, if we iteratively apply the above process based on the cases, then the process will end up that $F_1$ is isomorphic to $F_i$. (To here, we also ``expand" these combined labels in $F_1$ and $F_i$.)

The approximation algorithm Alg-{\sc app-uMaf} for {\sc app-uMaf} proceeds similarly with the Alg-{\sc app-rMaf} for {\sc app-rMaf}. Due to limit space, we will not present the details of Alg-{\sc app-uMaf} here. For {\sc app-uMaf}, we have the following theorem, whose proof is similar to that for Theorem~\ref{alg-app-rMaf-ratio}.

\begin{theorem}
\label{alg-app-uMaf-ratio}
Algorithm Alg-{\sc app-uMaf} is $4$-approximation algorithm for the {\sc app-uMaf} problem that runs in time $O(mn^2)$, where $n$ is the size of the label-set $X$ and $m$ is the number of forests in the input instance.
\end{theorem}

\section{Experiment}

We have implemented our algorithms Alg-{\sc para-rMaf} and Alg-{\sc app-rMaf} in C++, obtain programs {\sc Pmaf} and {\sc Amaf}, respectively.

Let $I=(F_1,F_2,\ldots,F_m)$ be an instance of {\sc Maf}. The program {\sc Pmaf} searches for the optimal value for the instance $I$ by starting with a lower bound $k$ of Opt$(I)$ and increasing $k$ until it can return an agreement forest for $I$ whose order is not greater than $k$. Firstly, we use the program {\sc Amaf} to get an agreement forest whose order is $k'$. According to the theoretical analysis given above, we have that $k'$ is not greater than $3$ times the optimal value for the instance. Therefore, $\lfloor k'/3 \rfloor \leq \mbox{Opt}(I)$, and the program {\sc Pmaf} can start with $\lfloor k'/3 \rfloor$.

We test our programs for both simulated and biological data on a 3.06Ghz Pentium(R) Dual-Core CPU system with 2GB of RAM running Windows XP.

\subsection{Simulated Data}

The simulated data are generated by using the following three-stage approach.

Firstly, generating a random rooted binary phylogenetic tree with $n$ labels. Our way of generating random rooted binary phylogenetic trees is the same as that in~\cite{29}. We use the integers from $1$ to $n$ to represent the $n$ irrelevant labels. At first, a bipartition on the $n$ integers is created by randomly cutting the list into two non-empty parts. This bipartition represents the edges adjacent to the root node of the tree being constructed. Then, each of the two induced partitions is randomly split into two lists to create a further bipartition of these sets. New bipartitions are then created recursively by cutting elements of previously created bipartitions into two sets until the bipartitions only consists of singleton elements. Thus, the tree is created by starting at the root and creating bipartitions (edges) until the leaf nodes are reached. The labels of the leaves are the singleton elements in the bipartitions, respectively. To maintain the consistency with the previous definition of phylogenetic trees, a new leaf labeled with $\rho$ would be attached to the root of the tree. And the leaf $\rho$ will be regarded as the new root of this tree.

Secondly, transforming the binary tree to a general tree. We randomly choose some internal edges in the binary tree, whose two ends are both internal vertices, to contract. The contracting operation is applied by removing the two endpoints $u$ and $v$ of the edge and introducing a new vertex which is adjacent to every vertex that is adjacent to at least one of $u$ and $v$. The number of edges that are contracted is also random.

Finally, transforming the original tree into other resulting trees by using a known number of SPR operations. Suppose $y$ resulting trees are constructed by applying $x$ SPR operations on the original tree $T_0$, respectively. Then the instance which consists of the original tree $T_0$ and the $y$ resulting trees $T_1,\ldots,T_{y}$ has an MAF whose order could not be greater than $x*y + 1$. Note that there are two reasons for that the order of the MAF for the instance could not be greater than $x*y + 1$: (1). the sequence of SPR operations we use to obtain $T_i$ from $T_0$, $1 \leq i \leq y$, may not be the shortest such sequence, that is, the order of the MAF for $T_0$ and $T_i$ maybe less than $x+1$; (2). the construction of $T_i$ and $T_j$ may use a same SPR operation, $1 \leq i \leq j \leq y$, that is, the order of the MAF for $T_i$ and $T_j$ maybe less than $2x+1$.

In the following discussion, we denote by t$n-m$ a set of instances of {\sc Maf}, each instance of which consists of $m$ phylogenetic trees with $n$-leaves.

\begin{figure}[h]
\begin{minipage}[t]{0.5\textwidth}
\centering
\includegraphics[width=3in]{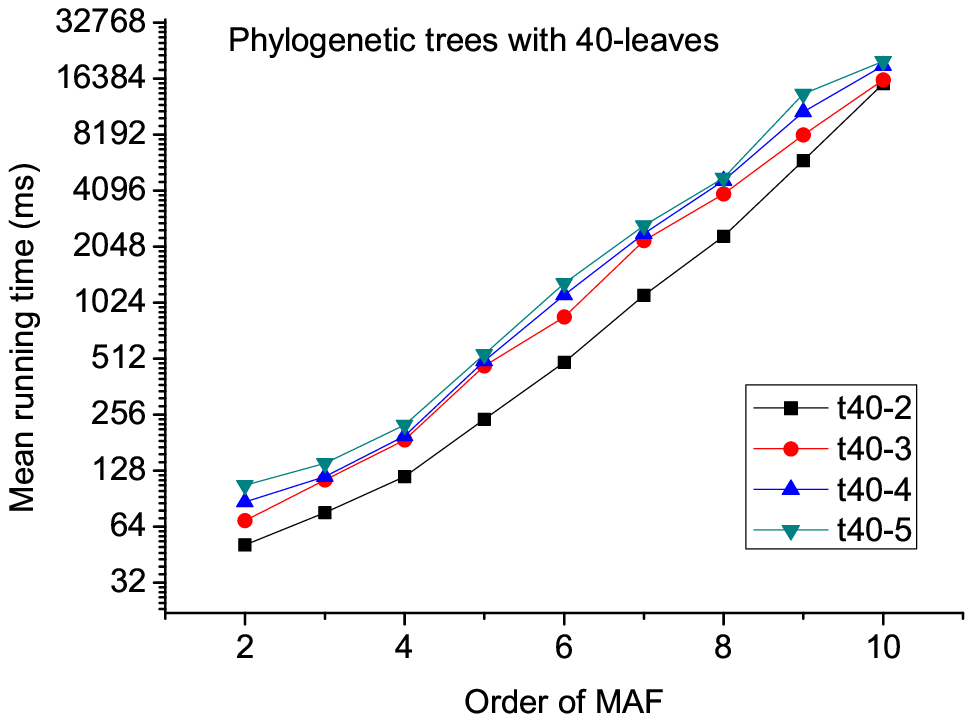}
\end{minipage}%
\begin{minipage}[t]{0.5\textwidth}
\centering
\includegraphics[width=3in]{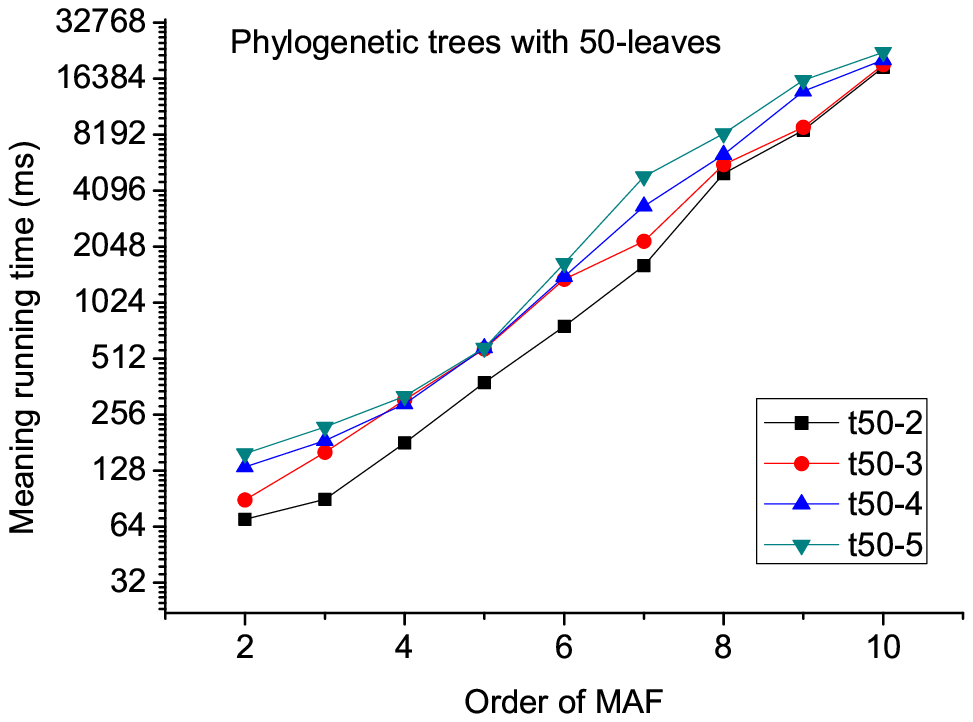}
\end{minipage}
\caption{Mean running time of {\sc Pmaf} on simulated data}
\label{average-time}
\end{figure}

We run our program {\sc Pmaf} on the simulated data set t$n-m$, where $n \in \{40,50\}$ and $m \in \{2,3,4,5\}$. For each instance set t$n-m$, $n \in \{40,50\}$, $m \in \{2,3,4,5\}$, it contains at least $20$ instances. Figure~\ref{average-time} shows the mean running time of {\sc Pmaf} on these simulated instances with the given order of MAF.  The slopes of the curves in the figure are always between $1$ and $\log_{2}3$, which indicating that the actual running time of the algorithm is between its worst-case running time of $O(3^k n^2m)$ and its best running time $O(2^k n^2m)$ (in the best case, all branches in the search tree of the algorithm for the instance always make two ways).

The running time of {\sc Amaf} on each instance of the simulated data set t$n-m$, where $n \in \{40,50\}$ and $m \in \{2,3,4,5\}$, is always less than one second.

\begin{figure}[h]
\begin{minipage}[t]{0.5\textwidth}
\centering
\includegraphics[width=3in]{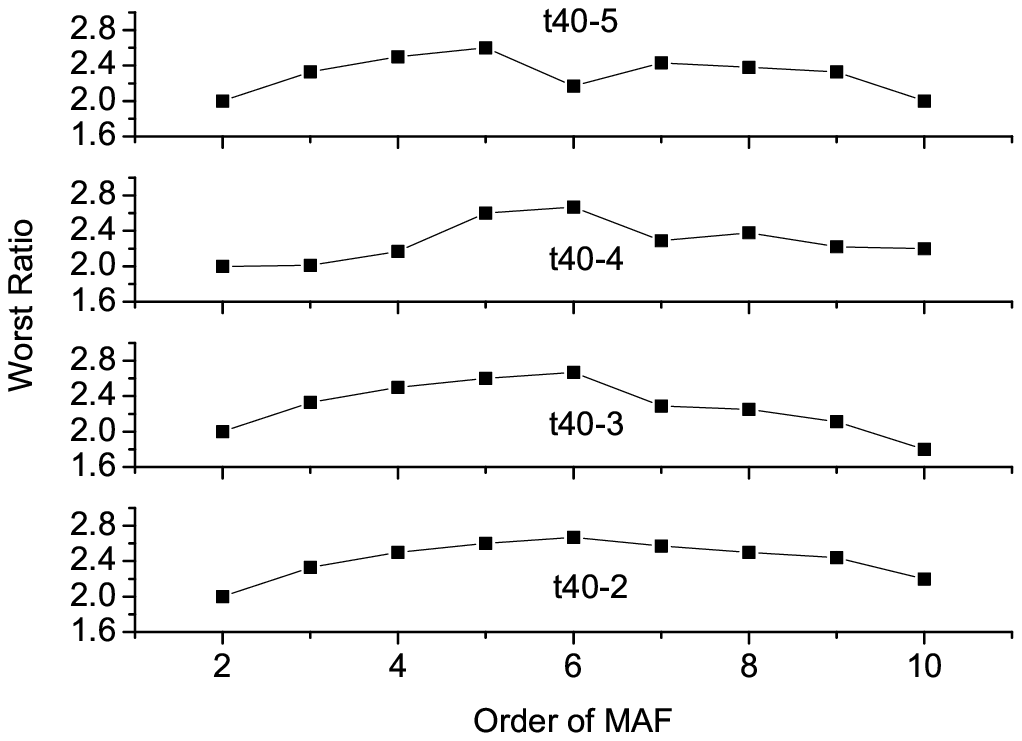}
\end{minipage}%
\begin{minipage}[t]{0.5\textwidth}
\centering
\includegraphics[width=3in]{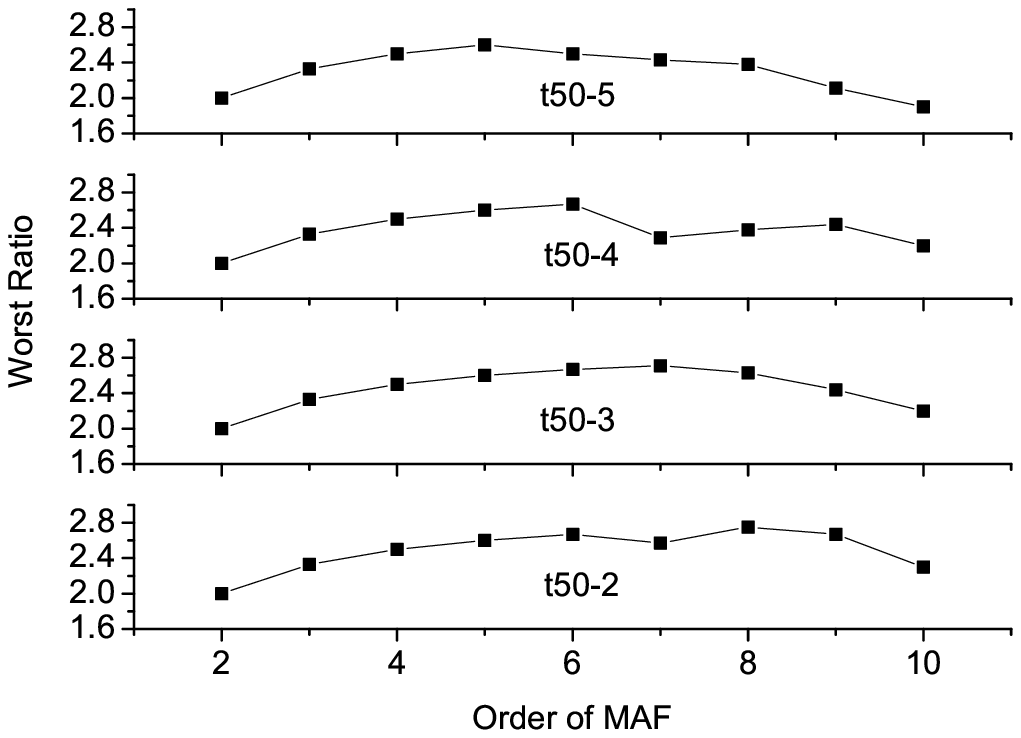}
\end{minipage}
\caption{Worst ratio of {\sc Amaf} on simulated data}
\label{worst-ratio}
\end{figure}

Figure~\ref{worst-ratio} shows the worst approximation ratio of {\sc Amaf} for the simulated data set t$n-m$, where $n \in \{40,50\}$ and $m \in \{2,3,4,5\}$. The top points of the curves in the figure are always not greater than $3$, which indicating that the actual approximation ratio of the algorithm is less than $3$. Moreover, as can be seen from the figure, when the order of MAF is greater than $6$, there is a down trending for the ratio of the algorithm with the order of MAF increasing. This is because as the increasing of the order of MAF, the number of ``right" edges in a tree, which are the edges that are not in the MAF for the instance, is increasing. Thus, the probability of removing ``right" edges by the algorithm is increasing. Therefore, the ratio of the algorithm trends down.

\subsection{Biological Data}

We run {\sc Pmaf} on the protein tree data set~\cite{29,30}. The protein tree data set consists of 22,437 binary protein trees, each constructed from a set of proteins covering from 4 to 144 microbial genomes. Among these trees, there are $15$ protein trees covering all 144 microbial genomes. Thus, the $15$ trees have the same label-set, and it is of biological meaning to compare these trees. We create extensive instances for the $15$ protein trees. The instances are created by the following way.

Firstly, create a random label-set $X'$ with fixed size not greater than $144$. Secondly, construct the subtrees with label-set $X'$ of the $15$ protein trees, respectively. Finally, randomly choose a fixed number of subtrees to compare.

\begin{figure}[h]
\begin{center}
\includegraphics[width=0.5\textwidth]{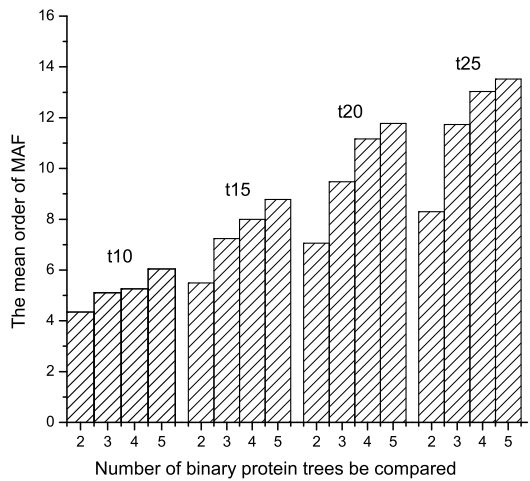}\\
\vspace*{-6mm}
\caption{Mean order of MAF for binary protein trees}
\label{mean-order}
\end{center}
\end{figure}

Figure~\ref{mean-order} shows the mean order of MAFs for the instances of protein trees. The number of instances we tested in each instance set t$n-m$ is not less than $50$, where $n \in \{10,15,20,25\}$ and $m \in \{2,3,4,5\}$. As can be seen from the Figure~\ref{mean-order}, we can get that the mean order of MAFs for the instances are increase as the number of trees in the instances increase, which indicates that the reticulation has influenced the evolutionary history of these parts of the genomes that being compared. Therefore, it makes perfect sense to study the {\sc Maf} problem on multiple trees.

We also run {\sc Pmaf} and {\sc Amaf} on the Poaceae data set from the Grass Phylogeny Working Group~\cite{31}. The dataset contains sequences for six loci: internal transcribed spacer of ribosomal DNA (ITS); NADH dehydrogenase, subunit F (ndhF); phytochrome B (phyB); ribulose 1,5-biphosphate carboxylase/oxygenase, large subunit (rbcL); RNA polymerase II, subunit $\beta''$ (rpoC2); and granule bound starch synthase I (waxy). When comparing these trees, only shared taxa of the set of trees are kept.

\begin{table}
\begin{center}
\scriptsize
  \begin{tabular}{|c|c|c|c|}
  \hline
  DataSet & $|X|$ & Order of MAF & Ratio\\
  \hline

  rpoC2, waxy, ITS & 10 & 5 & 1.6\\
  \hline

  ndhF, phyB, rbcL & 21 & 8 & 1.625\\
  \hline

  ndhF, phyB, rbcL, ropC2, ITS & 14 & 10 & 1.3\\
\hline

\end{tabular}
\vspace{0mm}
\caption{Comparing the trees of Poaceae data set.}
\label{Poaceae}
\end{center}
\end{table}

Tabel~\ref{Poaceae} shows our experimental results on the Poaceae data set. As can be seen from the table, the ratios outputted by {\sc Amaf} are small, because the size of instances are very small and the order of an MAF for these instances are large, relative to their sizes.

\begin{figure}[h]
\begin{center}
\includegraphics[width=0.5\textwidth]{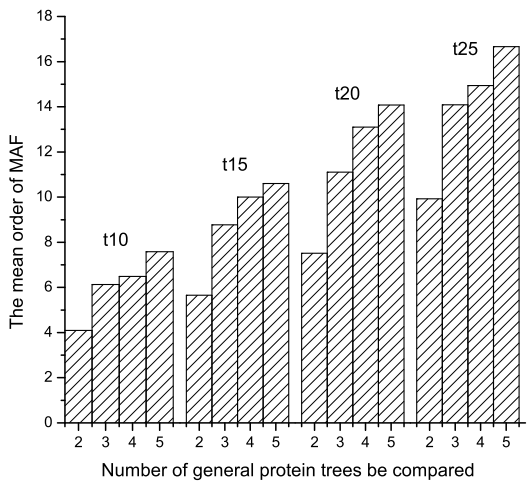}\\
\vspace*{-6mm}
\caption{Mean order of MAF for general protein trees}
\label{mean-nonbinary-order}
\end{center}
\end{figure}

The protein tree data set and Poaceae data set above, are all binary tree. In order to test our programs on biological data sets of general phylogenetic trees, we introduce the multifurcations in the $15$ protein trees which has been discussed above by collapsing each bipartition with a bootstrap support value of less than 0.8. The way for introduction of multifurcations is the same as the way in \cite{32}. Figure~\ref{mean-nonbinary-order} shows the mean order of MAFs for the instances of general protein trees. The number of instances we tested in each instance set t$n-m$ is not less than $50$, where $n \in \{10,15,20,25\}$ and $m \in \{2,3,4,5\}$.

\section{Conclusion}

In this paper, we presented the first group of parameterized algorithms for the Maximum Agreement Forest problem on multiple general phylogenetic trees: one for rooted trees that runs in time $O(3^k n^2m)$; and the other for unrooted trees that runs in time $O(4^k n^2m)$. We also presented the first group of approximation algorithms for the Maximum Agreement Forest problem on multiple general phylogenetic trees: one for rooted trees with ratio $3$, and the other for unrooted trees with ratio $4$.

Extensive experiments on simulated data and biological data show that our programs {\sc Pmaf} and {\sc Amaf} are able to calculate the  orders of MAFs for the constructed instances. In particular, when the order of MAF is small, {\sc Pmaf} can return the order of MAF for the instance quickly.

Further improvements on the algorithm complexity of parameterized algorithms and the ratio of polynomial-time approximation algorithms for the Maximum Agreement Forest problem on multiple general phylogenetic trees are certainly desired. However, the improvement seems require new observations in the graph structures of the phylogenetic trees.

\end{document}